\documentclass[epj]{svjour}
\usepackage{amsmath}
\usepackage{amssymb}

\usepackage{braket}            
\usepackage{graphicx}          
\usepackage[utf8]{inputenc}    
\usepackage{hyperref}          
\usepackage{color}             
\usepackage[caption=false]{subfig}        
\usepackage{csquotes}        
\usepackage{mathtools}         
\usepackage[normalem]{ulem}            
\usepackage{placeins}          
\usepackage[nointegrals]{wasysym}          
\usepackage{cite      }

\usepackage{bm}

\definecolor{darkgreen}{rgb}{0,0.45,0}

\newcommand{\rb}[1]{\left(#1\right)}
\newcommand{\sq}[1]{\left[#1\right]}
\newcommand{\expe}[1]{\text{e}^{#1}}

\newcommand{\mo}{\mathcal{O}}
\newcommand{\gfo}{\gamma_4}
\newcommand{\gfi}{\gamma_5}
\newcommand{\gmu}{\gamma_\mu}
\newcommand{\eabc}{\epsilon_{abc}}
\newcommand{\qo}[1]{q_{1,\,#1}}
\newcommand{\qto}[1]{q_{2,\,#1}}
\newcommand{\qth}[1]{q_{3,\,#1}}
\newcommand{\mcR}{\mathcal{R}}
\newcommand{\mpiv}{239(1)}

\newcommand{\Tpc}{T_{\rm pc}}
\newcommand{\bean}{\begin{eqnarray*}}
\newcommand{\eean}{\end{eqnarray*}}
\newcommand{\bea}{\begin{eqnarray}}
\newcommand{\eea}{\end{eqnarray}}
\newcommand{\be}{\begin{equation}}
\newcommand{\ee}{\end{equation}}
\newcommand{\nn}{\nonumber}
\newcommand{\tr}{\mbox{tr}\,}

\DeclareMathOperator\arccosh{arccosh}

\DeclareRobustCommand{\eqnr}[1]{Eq.~$\left(\ref{#1}\right)$}
\newcommand{\eqnrtwo}[2]{{Eqs.~$\rb{\ref{#1}}$ and $\rb{\ref{#2}}$}}

\newcommand{\Fig}[1]{Figure \ref{#1}}
\newcommand{\Tab}[1]{Table \ref{#1}}

\newcommand{\Sec}[1]{Section \ref{#1}}
\newcommand{\Figtwo}[2]{Figures \ref{#1} and \ref{#2}}

\newcommand{\Refl}[1]{Ref.~\cite{#1}}    

\begin{document}

\title{Non-zero temperature study of spin $1/2$ charmed baryons using lattice gauge theory }
\author{Gert Aarts\inst{1,2} \and Chris Allton\inst{1} \and M.~Naeem Anwar\inst{1} \and Ryan Bignell\inst{1} \and Timothy J.~Burns\inst{1} \and Benjamin J\"ager\inst{3} and Jon-Ivar Skullerud\inst{4}}

\institute{Department of Physics, Swansea University,\\Swansea, SA2 8PP, United Kingdom, \email{\{g.aarts,c.allton,m.n.anwar,t.burns\}@swansea.ac.uk,bignellr@tcd.ie}
\and
European Centre for Theoretical Studies in Nuclear Physics and Related Areas (ECT*) \& Fondazione Bruno Kessler, Strada delle Tabarelle 286, 38123 Villazzano (TN), Italy
\and
Quantum Field Theory Center \& Danish IAS, Department of Mathematics and Computer Science \\
University of Southern Denmark, 5230, Odense M, Denmark, \email{jaeger@imada.sdu.dk}
\and
Department of Theoretical Physics and Hamilton Institute, National University of Ireland Maynooth, 
County Kildare, Ireland \& School of Mathematics, Trinity College, Dublin, Ireland, \email{jonivar.skullerud@mu.ie}
}

\date{\today}

\abstract{
We study the behaviour of spin $1/2$ charmed baryons as the temperature increases. We make use of anisotropic lattice QCD simulations with $N_f = 2 + 1$ dynamical flavours. After determining the positive and negative parity ground state masses at the lowest temperature, we investigate the effect of rising temperature using ratios of thermal lattice correlators with both so-called reconstructed correlators and with simple model correlators. This avoids difficulties associated with non-zero temperature fitting or spectral reconstruction. We find that temperature effects are prominent throughout the hadronic phase for all negative parity channels considered and for some positive parity channels. Subsequently and where possible, we determine the masses of the ground states as a function of temperature. Finally we consider the effect of chiral symmetry restoration and extract an estimate of the pseudocritical temperature from singly charmed baryonic correlators.
}

\PACS{
{14.20.Lq}{charmed baryons}
\and
{11.15.Ha}{Lattice gauge theory}
\and
{12.38.Mh}{Plasmas, quark-gluon}
}

\maketitle


\section{Introduction}
\label{sec:intro}

This work focuses on the fate of spin $1/2$ charmed baryons as the temperature increases and the confining hadronic medium at low temperature smoothly transitions to a quark-gluon plasma (QGP), as predicted by Quantum Chromodynamics (QCD)~\cite{Aoki:2006we}. While the light degrees of freedom deconfine at the crossover \cite{Borsanyi:2010cj,Bazavov:2011nk}, there is evidence that heavier hadrons comprised of charm and bottom quarks can survive as bound states in the QGP. Since the seminal work by Matsui and Satz \cite{Matsui:1986dk}, this has been studied extensively for charmonium ($c\bar c$) and bottomonium ($b\bar b$) states, as it provides important insights into the length scales in the QGP~\cite{Burnier:2007qm,Brambilla:2010vq,Strickland:2011aa}. For results obtained with lattice QCD, see e.g.\ the comprehensive reviews~\cite{Rothkopf:2019ipj,Guenther:2020jwe} and references therein.

Here we consider baryons containing charm quarks,  building upon previous work studying the behaviour of light and strange baryons at non-zero temperature~\cite{Aarts:2015mma,Aarts:2017rrl,Aarts:2018glk}. The focus in this earlier work was parity doubling, the emergence of a degeneracy between positive- and negative-parity correlators and ground state masses, which is a signal for chiral symmetry restoration. Due to the heavier mass of the charm quark, one expects a reduction in the level of parity doubling observed when compared to the light and strange baryons. Instead, baryons containing multiple charm quarks may be more similar to heavy-quark mesons (quarkonia) and therefore not immediately dissolve in the QGP. Hence, it will be interesting to analyse the structure of charmed baryons across a range of temperatures.

Heavy-light charmed $D$ and $D_s$ mesons have been studied using thermal lattice QCD  in Refs.~\cite{Kelly:2018hsi, Aarts:2022krz}. 
In particular, \Refl{Aarts:2022krz} considered the response of $D_{\rb{s}}$ mesons to an increasing temperature without the explicit need for non-zero temperature fits and spectral reconstruction. 
The outcome of this analysis was subsequently used to inform the applicability of standard fits to the correlation function used to determine the ground state mass. 
The analysis studied ratios of lattice correlators to single-state model correlators to determine when the spectral content of the correlator was different from that of the zero-temperature correlator. Here we extend this treatment to the baryonic sector for the first time.

Motivated and directed by the outcome of the analysis of correlator ratios across temperatures, we fit correlators with standard exponential fits when ratios suggest that thermal effects are small. Hence these fits assume that the hadron is a (narrow) bound state. Extensive use is made of model averaging methods~\cite{Jay:2020jkz,Rinaldi:2019thf} in order to produce a reliable extraction of the mass. It is noted that the behaviour of charmed baryon masses at non-zero temperature may be of phenomenological interest~\cite{Yao:2018zze}.

To make use of the full range of temperatures available, we also compare positive and negative parity channels using the correlation functions directly~\cite{Aarts:2017rrl,Aarts:2018glk,Aarts:2020vyb}, to reveal the effect of chiral symmetry restoration in the QGP. Parity doubling is not expected due to the large charm quark mass; nevertheless, for singly charmed baryons a clear change of behaviour is observed.

The paper is organised as follows. In \Sec{sec:LQCD} we summarise the FASTSUM Generation 2L ensembles, correlator methods, and the fitting techniques used. We consider ratios of the lattice correlator with the \enquote{reconstructed} correlator  and with a single-state model correlator in \Sec{sec:ratios}. In \Sec{sec:masses} we determine masses across a range of temperatures provided the analysis of the previous section suggests this is appropriate. The fit-independent parity doubling ratio is presented in \Sec{sec:par}. 
Conclusions and future work are discussed at the end.

\section{Lattice Correlators}
\label{sec:LQCD}

In the following subsections we discuss the gauge field ensembles, operators, and fitting procedure used to determine ground state masses at the lowest temperature.

\subsection{Ensembles}
\label{sec:LQCD:Ens}

\begin{table}[b]
  \centering
    \caption{\textsc{FASTSUM} Generation 2L ensembles used in this work. The lattice size is $32^3 \times N_\tau$, with temperature $T = 1/\rb{a_\tau N_\tau}$. The spatial lattice spacing is $a_s = 0.11208\rb{31}$ fm, renormalised anisotropy $\xi = a_s/a_\tau = 3.453(6)$ and the pion mass $m_\pi = \mpiv$ MeV~\cite{Wilson:2019wfr}. We use $\sim 1000$ configurations and eight (random) sources for a total of $\sim 8000$ measurements at each temperature. The estimate for $\Tpc$ comes from an analysis of the renormalised chiral condensate and equals $\Tpc = 167(2)(1)$ MeV~\cite{Aarts:2020vyb,Aarts:2022krz}. Full details of these ensembles may be found in Refs.~\cite{Aarts:2020vyb,Aarts:2022krz}.
    }
  \begin{tabular}{r|rrrrrr}
  $N_\tau$ & 128 & 64 & 56 & 48 & 40  & \\ \hline
  $T\,\rb{\text{MeV}}$ & 47 & 95 & 109 & 127 & 152 & \\
   $N_\tau$ & 36 & 32 & 28 & 24 & 20 & 16\\ \hline
  $T\,\rb{\text{MeV}}$ & 169 & 190 & 217 & 253 & 304 & 380 \\
  \end{tabular}
\label{tab:ensembles}
\end{table}

We make use of the thermal ensembles of the \textsc{FASTSUM} collaboration~\cite{Aarts:2020vyb}, with $2+1$ flavours of Wilson fermions on  anisotropic lattices. The renormalised anisotropy is $\xi \equiv a_s/a_\tau = 3.453(6)$ \cite{Dudek:2012gj,Aarts:2020vyb}. The lattice action follows that of the Hadron Spectrum Collaboration~\cite{Edwards:2008ja} and is a Symanzik-improved~\cite{Symanzik:1983dc,Symanzik:1983gh} anisotropic gauge action with tree-level mean-field coefficients and a mean-field-improved Wilson-clover~\cite{Sheikholeslami:1985ij,Zanotti:2004qn} fermion action with stout-smeared links~\cite{Morningstar:2003gk}. Full details of the action and parameter values can be found in \Refl{Aarts:2020vyb}. We use the \enquote{Generation 2L} ensembles which have a pion mass of $m_\pi = \mpiv$ MeV (in the previous \enquote{Generation 2} ensembles the pion mass was $m_\pi = 384(4)$ MeV~\cite{HadronSpectrum:2008xlg,Aarts:2014nba,Aarts:2014cda}). While this is still heavier than physical, it represents an important step towards the physical regime. The strange quark has been tuned to its physical value via the tuning of the light and strange pseudoscalar masses~\cite{HadronSpectrum:2008xlg,HadronSpectrum:2012gic,Cheung:2016bym}.
The ensembles are generated using a fixed-scale approach, such that the temperature is varied by changing $N_\tau$, as $T=1/\rb{a_\tau N_\tau}$. A summary of the ensembles is given in \Tab{tab:ensembles}. There are five ensembles below the pseudocritical temperature $\Tpc = 167(2)(1)$ MeV, one close to $\Tpc$ and five above $\Tpc$. The estimate for $\Tpc$ comes from an analysis of the renormalised chiral condensate~\cite{Aarts:2020vyb}. Note that here we have used the updated lattice spacing of \Refl{Wilson:2019wfr}, which has been implemented in our analysis in Ref.~\cite{Aarts:2022krz}.

\subsection{Operators}
\label{sec:LQCD:Op}

The $J=1/2$ singly and doubly charmed baryons, with quark content, are grouped according to the underlying flavour symmetry of QCD, namely
\begin{align}
SU(3)~\mathbf{\bar{3}}:& \quad \Lambda_{c}\rb{udc}, ~ \Xi_{c}\rb{usc} \nonumber \\
SU(3)~\mathbf{6} : & \quad \Sigma_{c}\rb{udc}, ~ \Xi^\prime_{c}\rb{usc},~\Omega_{c}\rb{ssc}
\end{align}
and for doubly charmed baryons, 
\be
SU(3) \times U(1)_\text{charm}~\mathbf{20}_M: \quad \Xi_{cc}\rb{ccu},  ~ \Omega_{cc}\rb{ccs},
\ee
following the decomposition of $SU(4)$ as in the PDG 2020~\cite{ParticleDataGroup:2020ssz}.

The baryon correlators we consider are of the form
\begin{align}
  G^{\alpha\alpha^\prime}(x) = \Braket{\mo^\alpha(x)\overline{\mo}^{\alpha^\prime}(0)}
\end{align}
where $\overline{\mo} = \mo^\dagger\gfo$ and $\alpha$, $\alpha^\prime$ are Dirac indices.
The operators used follow Refs.~\cite{Leinweber:2004it,Edwards:2004sx}. Denoting the three quarks as $q_1$, $q_2$ and $q_3$ (from left to right), the operator for the case $q_1=q_2$ ($\Xi_{cc}$, $\Omega_{c}$ and $\Omega_{cc}$) is given by
\be
  \mo^{\alpha}_{{\rm 2fl}, 1/2}\rb{q_1,q_1,q_3} = \eabc\,\qo{\alpha}^{a}\rb{\qo{\beta}^b\sq{C\gfi}_{\beta\gamma}\qth{\gamma}^{c}}.\label{eqn:OP2fl12}
\ee
For the non-degenerate case, $\Sigma_c$ and $\Xi^\prime_c$ belong to the $SU(3)\,\, \mathbf{6}$ flavour multiplet, with operators
\begin{align}
\nn
  \mo^\alpha_{\mathbf{6}, 1/2}\rb{q_1,q_2,q_3} = & \frac{1}{\sqrt{2}}\eabc 
  \left(\qo{\alpha}^{a}\rb{\qth{\beta}^b\sq{C\gfi}_{\beta\gamma}\qto{\gamma}^{c}}\right.
\\ 
& + \left.\qto{\alpha}^{a}\rb{\qth{\beta}^b\sq{C\gfi}_{\beta\gamma}\qo{\gamma}^{c}}\right), 
\end{align}
  where $q_1=u$, $q_3=c$ and $q_2=d/s$ respectively. $\Lambda_{c}$ and $\Xi_{c}$ enjoy $SU(3) \,\,\mathbf{\overline{3}}$ flavour symmetry, with operators
\begin{align}
 \mo^\alpha_{\mathbf{\bar 3}, 1/2}\rb{q_1,q_2,q_3} 
 = & \frac{1}{\sqrt{6}}\eabc \Big(2\qth{\alpha}^{a}\rb{\qo{\beta}^{b}\sq{C\gfi}_{\beta\gamma}\qto{\gamma}^c}
 \nn \\ 
 & + \qto{\alpha}^{a}\rb{\qo{\beta}^{b}\sq{C\gfi}_{\beta\gamma}\qth{\gamma}^{c}} 
 \nn \\ 
 & - \qo{\alpha}^{a}\rb{\qto{\beta}^{b}\sq{C\gfi}_{\beta\gamma}\qth{\gamma}^{c}}\Big), 
\end{align}
where $q_1=u$, $q_3=c$ and $q_2=d/s$ respectively. 
The Euclidean gamma matrices satisfy $\gmu^\dagger = \gmu = \gmu^{-1}$ for $\mu=1,\dots4$, $\gfi = \gfi^\dagger = \gamma_{1}\gamma_{2}\gamma_{3}\gamma_{4}$ and $C$ is the charge conjugation matrix~\cite{Gattringer:2010zz}.

\begin{figure*}[t]
  \includegraphics[width=2.0\columnwidth]{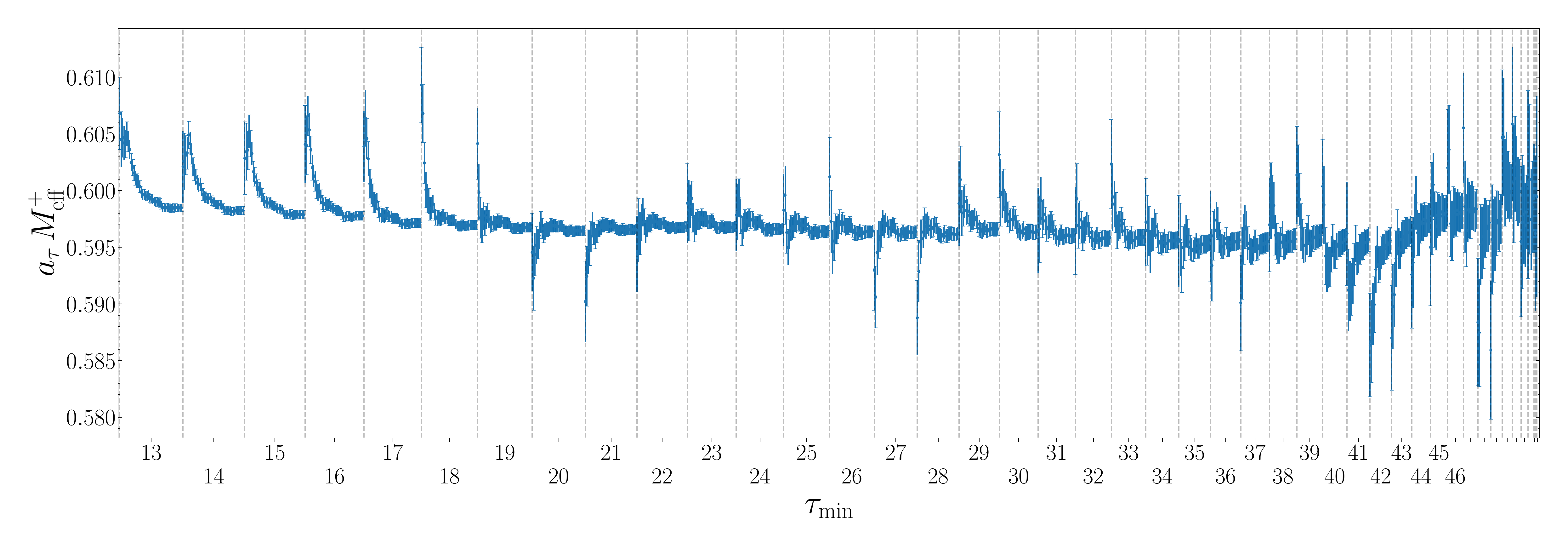}
  \caption{\label{fig:platAll}Mass of the positive parity $\Xi_{cc}$ ground state at the lowest temperature, from forward constant fits starting at $\tau_{\rm min}$ to the effective mass. For each $\tau_{\rm min}$, $\tau_{\rm max}$ increases from left to right in each pane, i.e.\ $\tau_{\rm min}<\tau_{\rm max}\leq N_\tau / 2 - 0.05\,N_\tau$; hence the \enquote{middle} $10$\% of points are excluded.
  X-axis labels are not shown where they would overlap, i.e.\ on the right side of the plot.}
\end{figure*}

The contractions of these operators produce terms proportional to
\begin{align}
      S_1 \tr\sq{C\Gamma_{\rm so}S_2C\Gamma_{\rm si}S_3}
  \quad \mbox{and} \quad
    S_1C\Gamma_{\rm so}S_2C\Gamma_{\rm si}S_3,
\end{align}
where $S_{1,2,3}$ is a quark propagator of flavour $q_{1,2,3}$ and $\Gamma_{\rm so/si}$ is the gamma matrix in the source/sink operator respectively. These are the principal building blocks of the baryon correlation functions.

The positive and negative parity projectors are~\cite{Gattringer:2010zz,Leinweber:2004it} 
\begin{align}
  P_\pm = \frac{1}{2}\left(1 \pm \gamma_4\right),
\end{align}
and we denote the projected correlation functions at vanishing spatial momentum as \cite{Aarts:2015mma,Aarts:2017rrl,Aarts:2018glk}
\be
G_\pm(\tau) = \tr P_\pm G(\tau).
\ee
These are related as \cite{Aarts:2017rrl}
\be
\label{eq:Gpm}
G_\pm(\tau) = -G_\mp(1/T -\tau),
\ee
implying that the forward- (backward-) propagating states of $G_+(\tau)$ are states with positive (negative) parity.

Gaussian smearing is applied to the source and sink using~\cite{Gusken:1989qx}
\begin{align}
  \eta^\prime = C_{\rm norm}\rb{1 + \kappa H}^n\eta,
\end{align}
where $\eta$ is the bare (delta function) source, $\kappa$ and $n$ determine the amount of smearing, $H$ is the spatial hopping part of the Dirac operator and $C_{\rm norm}$ is an appropriate normalisation. Here we used $\kappa = 5.5$, $n = 100$ at all temperatures. The root-mean-square radius of this profile is $\sim 6.8$ lattice sites. These parameters were chosen such that the positive parity nucleon ground state at the lowest temperature displays good ground state isolation, as in Refs.~\cite{Aarts:2017rrl,Aarts:2018glk}.

\subsection{Mass extraction}
\label{sec:Mass}

A systematic extraction of hadron masses from lattice QCD is an area of active development, with methods becoming increasingly sophisticated~\cite{Michael:1985ne,Blossier:2009kd,Bouchard:2014ypa,Alexandrou:2014mka,Stokes:2018emx,Rinaldi:2019thf,NPLQCD:2020ozd,Jay:2020jkz,Batelaan:2022fdq}. Broadly these can be divided into methods which improve the underlying correlation function~\cite{Michael:1985ne,Blossier:2009kd,Stokes:2018emx} and those which aim to more reliably extract the mass from a given correlation function. As mentioned above, we use gauge invariant Gaussian source and sink smearing in order to improve the overlap of the operator with the (zero-temperature) ground state. We also adopt the model averaging methods of both Refs.~\cite{Rinaldi:2019thf,Batelaan:2022fdq} and \Refl{Jay:2020jkz}. By comparing these two distinct methods we ensure a robust determination of the mass.

We start at the lowest temperature ($T=47$ MeV, $N_\tau=128$) and consider fit functions of the form
\begin{align}
  \overline{G}\rb{\tau} = \sum_{n=1}^{N}\, A_n\expe{-a_\tau M_n^{+}\tau/a_\tau} + B_n\expe{-a_\tau M_n^{-}\rb{N_\tau - \tau/a_\tau}},
  \label{eqn:expFit}
\end{align}
where $\overline{G}(\tau) = G_+\rb{\tau}/G_+\rb{0}$ and $M_n^{\pm}$ is the $n^{\rm th}$ positive (negative) parity state. The number of exponentials $N$ is allowed to vary from one to three. The final number of exponentials is set by examining the Gaussian Bayes factor and stepping back one exponential when it ceases to change as more exponentials are added~\cite{Lepage:2001ym,Hornbostel:2011hu,Bouchard:2014ypa,peter_lepage_2021_5777652}. This is equivalent to stopping when the data can no longer support additional exponential terms. 
We illustrate some of our findings below using the positive and negative parity $\Xi_{cc}$ ground states as examples.

\begin{figure*}[tb]
  \includegraphics[width=2.0\columnwidth]{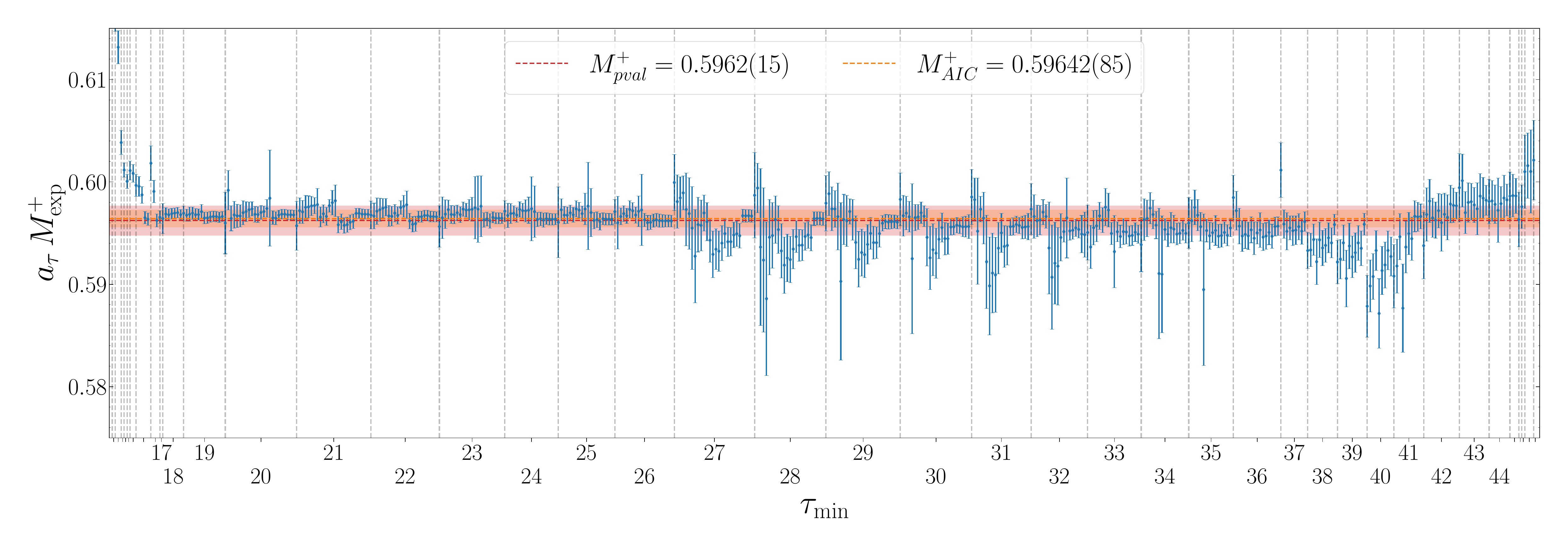}%
  \caption{\label{fig:expFits}Mass of the positive parity $\Xi_{cc}$ ground state at the lowest temperature, extracted from multi-exponential fits as in \eqnr{eqn:expFit}. For each $\tau_{\rm min}$, $\tau_{\rm max}$ increases from left to right in each pane. The top $30\%$ highest weighted fits from \eqnr{eqn:AIC} are shown, as are the corresponding model averaged values from \eqnrtwo{eqn:pVal}{eqn:AIC}. X-axis tick labels are not shown where they would overlap.}
\end{figure*}

Typically the correlator $\overline{G}\rb{\tau}$ is considered from a starting time $\tau_{\rm min}$ up until an end time $\tau_{\rm max}$. This is due to excited state contamination at early times and the onset of noise of late times (on the positive parity side and trivially reversed for the negative parity side).
The choice of \enquote{fit window} can have a substantial effect on the extracted mass; this is particularly evident for the effective mass~\cite{Gattringer:2010zz} fits shown in \Fig{fig:platAll} for the positive parity $\Xi_{cc}$ ground state at the lowest temperature.
Here we have fit the effective mass, 
\begin{align}
M_{\rm eff} = \frac{1}{\delta_\tau}\,\arccosh{\frac{\overline{G}\rb{\tau + \delta_\tau} + \overline{G}\rb{\tau - \delta_\tau}}{2\,\overline{G}\rb{\tau}}},
\end{align}
with $\delta_\tau = 2$ across an interval $\tau_{\rm min}\leq \tau\leq \tau_{\rm max}$, with varying $\tau_{\rm min}$. Specifically, we used $\tau_{\rm min} \geq 0.1\,N_\tau = 13$ and $\tau_{\rm max} \leq N_\tau / 2 - 0.05\,N_\tau = 57$ 
(note that in this section we use ``temporal lattice units'' for both Euclidean time and quoted mass values). 
This range was chosen to remove early time slices dominated by excited state effects and \enquote{noisy} time slices near the middle of the time extent. At small values of $\tau_{\rm max}$ clear excited state contamination is manifest in the increased value from the fit. As $\tau_{\rm min}$ and $\tau_{\rm max}$ are increased, excited state contamination is decreased as the ground state becomes more dominant, but the statistical uncertainty also increases.  While it may be possible to extract a robust ground state mass from the effective mass for $N_\tau=128$, in general this will not be possible at higher temperatures, due to a reduction in the number of temporal points.

Exponential fits electing a single fit window ignore the information from other fit windows. It is therefore beneficial to use model averaging methods to determine weighted fit parameters. 
The first method, introduced in \Refl{Rinaldi:2019thf} uses a weight
\begin{align}
  \tilde{w}_f^{(1)} = \frac{p_f/\rb{\delta M_f}^{2}}{\sum_{f^\prime=1}^{N} p_{f^\prime}/\rb{\delta M_{f^{\prime}}}^{2}},
  \label{eqn:pVal}
\end{align}
where $p_f$ is the p-value of the fit $f$, $\delta M_{f}$ is the uncertainty in the fit parameter of fit $f$ and there are $N$ fits to be averaged. This method penalises both poor fits and unconstraining fits~\cite{NPLQCD:2020ozd}.

This is in contrast to the second method~\cite{Jay:2020jkz} which uses an exponential of the modified Akaike information criterion~\cite{akaike1998bayesian,Jay:2020jkz}
\begin{align}
  \tilde{w}_{f}^{(2)} = \exp\rb{-\frac{1}{2}\chi^2_{\rm aug}\rb{M_f}- k- N_{\text{cut}}},
  \label{eqn:AIC}
\end{align}
where $N_{\text{cut}}$ is the number of data points not fit to, $k$ is the number of fit parameters and $\chi^2_{\rm aug}$ is the augmented chi-squared~\cite{Lepage:2001ym}. 
Agreement between the two methods provides confidence in the extracted fit parameter.

A large number of roughly equivalent fit windows is shown in \Fig{fig:expFits}. The exponential fits have range $\tau_{\rm min}=1\le \tau \le \tau_{\rm max}\leq N_\tau /2 - 0.05\,N_\tau$. In particular, in \Fig{fig:expFits} only the top $30\%$ highest weighted fits using the weight of \eqnr{eqn:AIC} are shown.

\begin{figure}[tb]
  \includegraphics[width=\columnwidth]{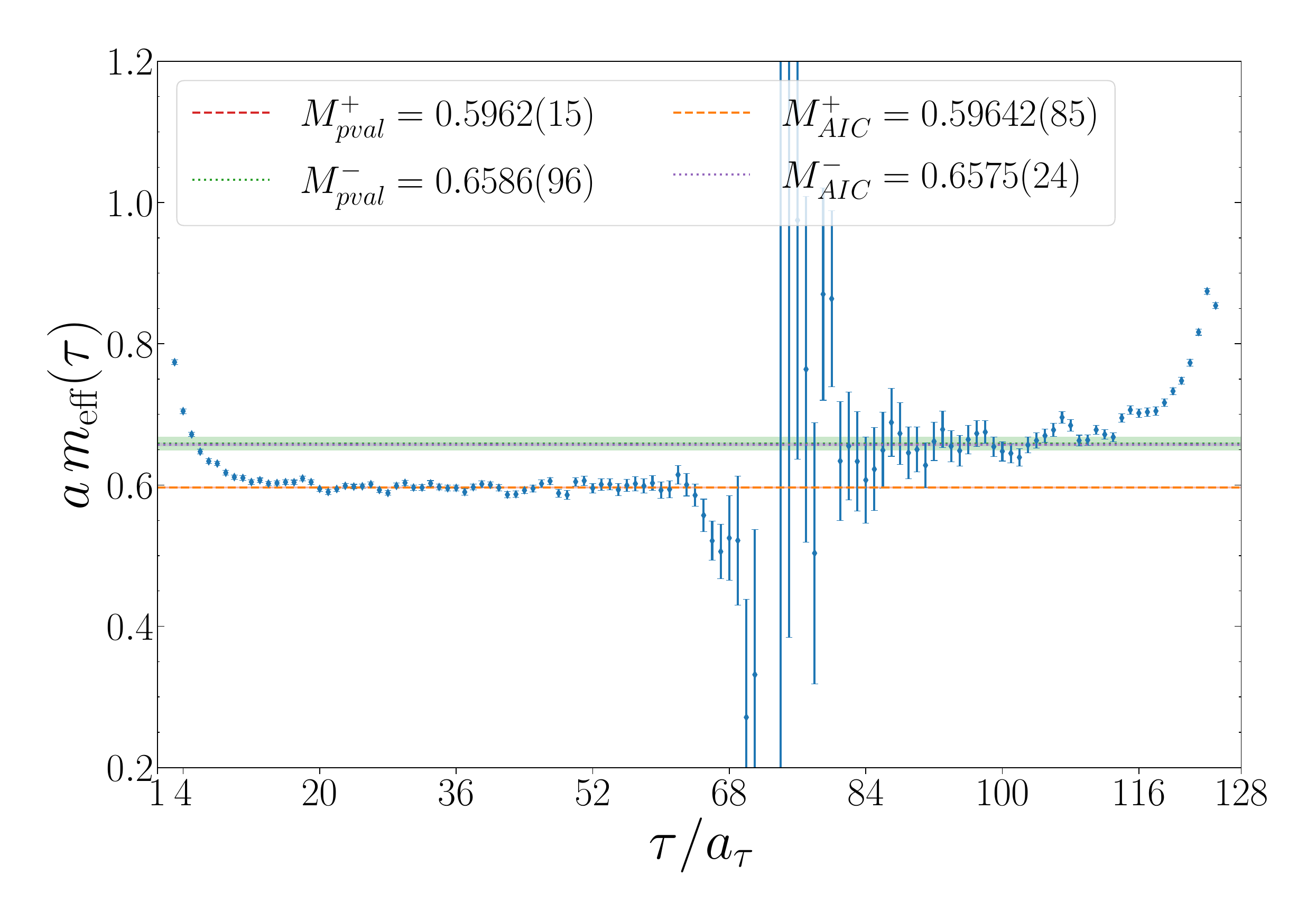}
  \caption{\label{fig:effMass}Comparison of the effective mass for the positive and negative parity $\Xi_{cc}$ ground state at the lowest temperature with the results of the two model averaging methods of \eqnrtwo{eqn:pVal}{eqn:AIC}.
  }
\end{figure}

\begin{figure}[!p]
  \includegraphics[width=\columnwidth]{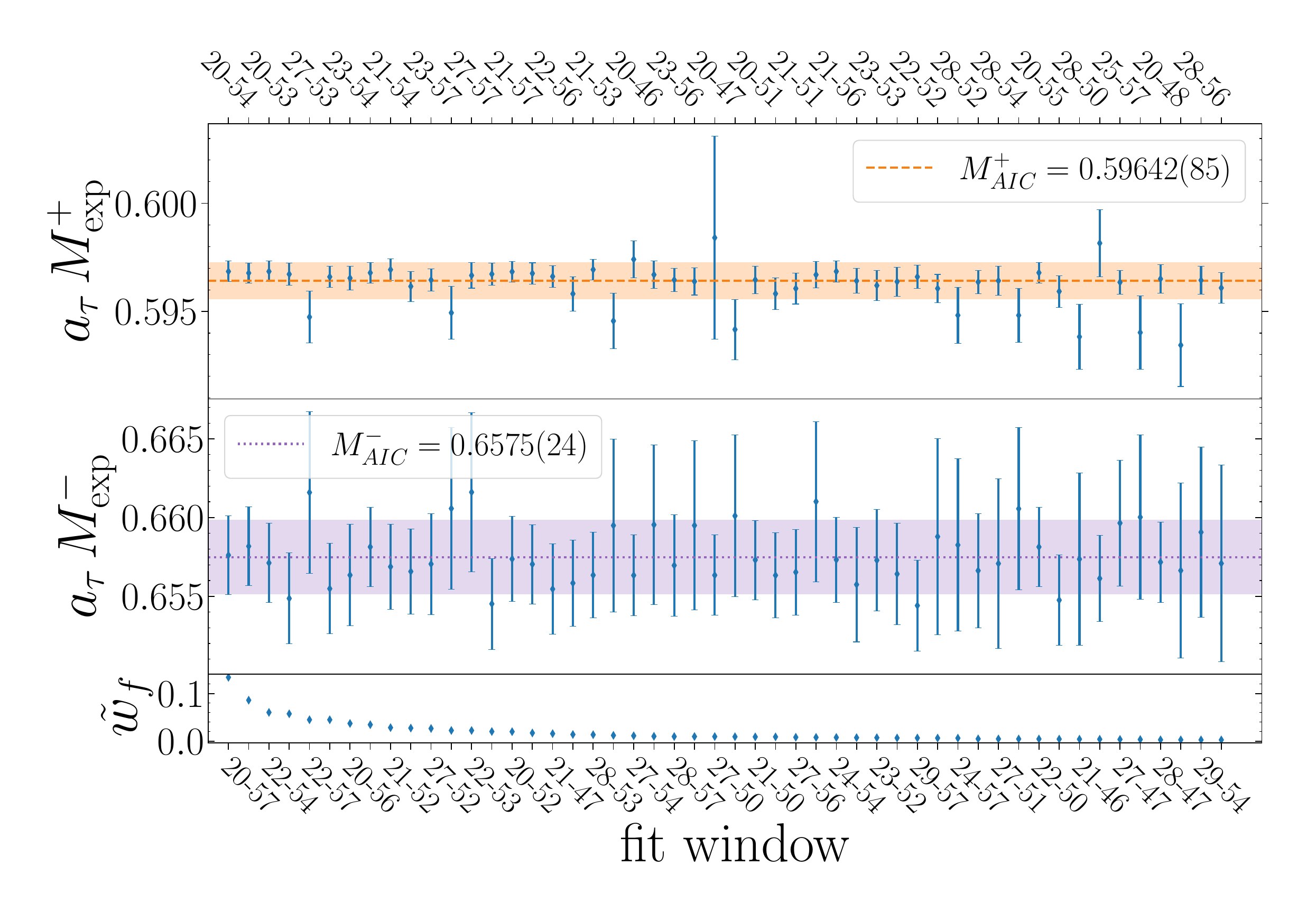}
  \caption{\label{fig:Weights}Highest weighted fit results for 
  the positive and negative parity $\Xi_{cc}$ ground state at the lowest temperature as a function of fit window. Both sides of the correlator are fit symmetrically and so e.g. \enquote{20-57} fits points $\left[20,\,57\right]$ and $\left[71,\,108\right]$ simultaneously. The weights $\tilde{w}_{f}^{(2)}$ of \eqnr{eqn:AIC} are also shown, as is the resulting model averaged fit value.}
\vspace*{0.4cm}
  \includegraphics[width=\columnwidth]{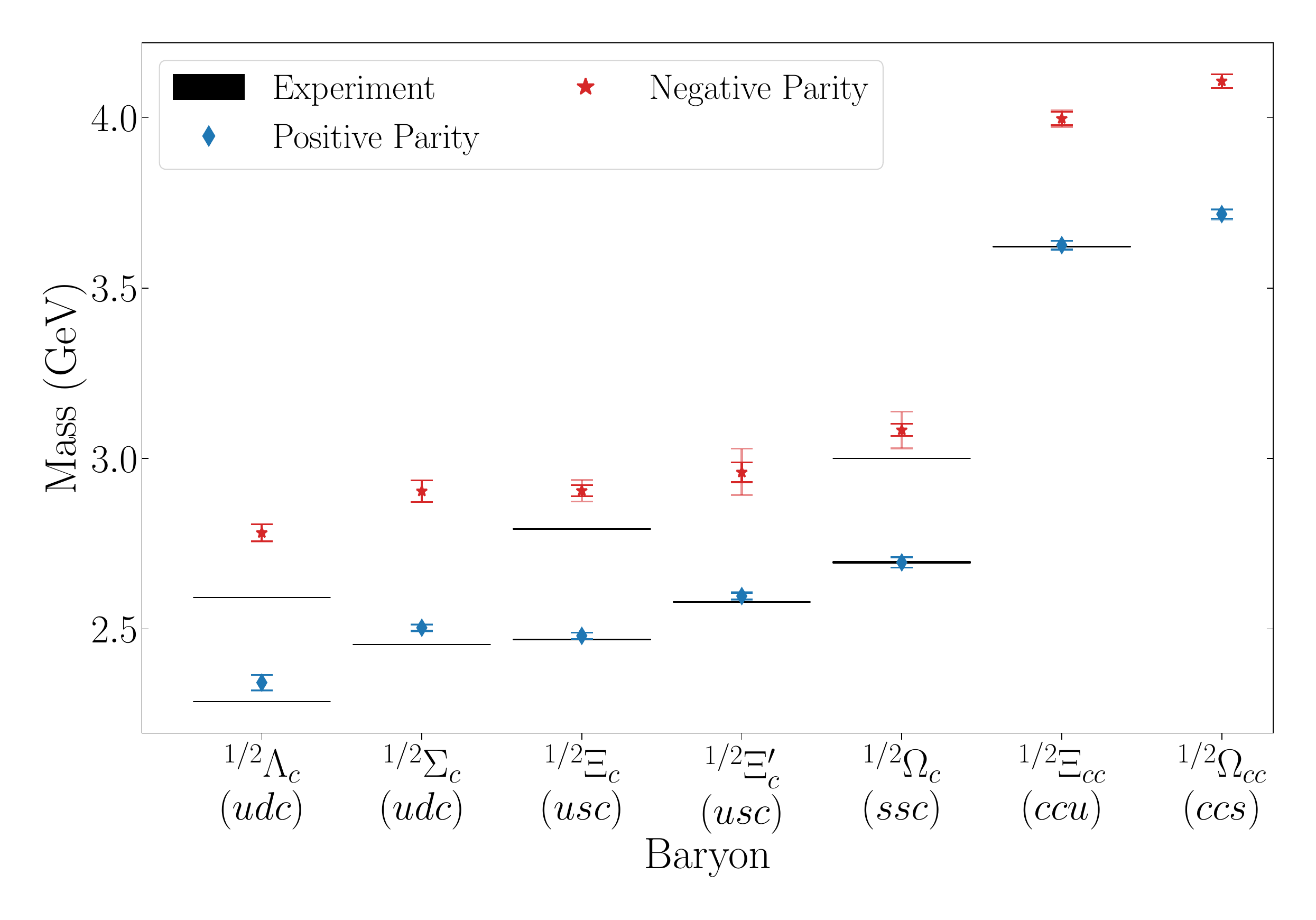}
  \caption{\label{fig:spectAll}Positive (diamonds $\lozenge$) and negative (stars $\star$) parity ground state masses $M^\pm$ at the lowest temperature. Inner error bars represent the statistical uncertainty and the outer bars incorporate a systematic error from the choice of averaging method. The grey bars represent experimental results from the Particle Data Group, when known~\cite{Workman:2022ynf}.}
\end{figure}

To exclude outlying fits, a further cut is made. The top $30\%$ of weighted fits using the modified Akaike information criterion weights of \eqnr{eqn:AIC} are separated and re-averaged using both methods. This is demonstrated in \Fig{fig:effMass}, in which the effective masses for both the positive and negative parity $\Xi_{cc}$ ground state are shown and compared to the model averaged results from the multi-exponential fits after this cut has been applied. The resulting fits are in good agreement for both the sectors giving us assurance that the fitted masses are robustly determined.

\begin{table*}[t]
\caption{Positive and negative parity ground state masses $M^\pm$ at the lowest temperature, in units of $a_\tau$ and in GeV. The systematic error due to the choice of model averaging method has been added in quadrature to the error from the selected model averaging procedure. The final two columns show the Particle Data Group values \cite{Workman:2022ynf}, when known, averaged over charge partners with systematic uncertainties added in quadrature.}
\label{tab:NT128-masses}
\centering
\begin{tabular}{l|llll|ll}
 & \multicolumn{4}{c|}{Lattice} & \multicolumn{2}{c}{PDG} \\ 
 Hadron & $a_\tau M^{+}$ & \multicolumn{1}{l|}{$a_\tau M^{-}$} & $M^{+}$ [GeV] & $M^{-}$ [GeV] & $M^{+}$ [GeV] & $M^{-}$ [GeV] \\ \hline
$\Lambda_{c}\rb{udc}$ & 0.3853(36) & \multicolumn{1}{l|}{0.4576(39)} & 2.342(22) & 2.782(24) & 2.28646(14) & 2.59225(28) \\
$\Sigma_{c}\rb{udc}$ & 0.4118(11) & \multicolumn{1}{l|}{0.4777(50)} & 2.5034(86) & 2.904(31) & 2.45379(30) & \multicolumn{1}{c}{$-$} \\
$\Xi_{c}\rb{usc}$ & 0.40787(89) & \multicolumn{1}{l|}{0.4779(49)} & 2.4796(77) & 2.905(30) & 2.46907(36) & 2.79290(71) \\
$\Xi^\prime_{c}\rb{usc}$ & 0.4271(15) & \multicolumn{1}{l|}{0.487(11)} & 2.596(11) & 2.961(67) & 2.57845(71) & \multicolumn{1}{c}{$-$} \\
$\Omega_{c}\rb{ssc}$ & 0.4433(21) & \multicolumn{1}{l|}{0.5072(87)} & 2.695(14) & 3.083(53) & 2.6952(17) & 3.00041(22) \\
$\Xi_{cc}\rb{ccu}$ & 0.59642(87) & \multicolumn{1}{l|}{0.6575(34)} & 3.6258(95) & 3.997(22) & 3.62160(40) & \multicolumn{1}{c}{$-$} \\
$\Omega_{cc}\rb{ccs}$ & 0.6114(17) & \multicolumn{1}{l|}{0.6757(26)} & 3.717(13) & 4.107(18) & \multicolumn{1}{c}{$-$} & \multicolumn{1}{c}{$-$}
\end{tabular}
\end{table*}

In practice, to produce a single number for a mass, a single averaged value is chosen by examination of the weights, such that the averaged value clearly represents the fits. This can be seen in \Fig{fig:Weights} where the averaged value encompasses the majority of the fit weights (which sum to 1). The difference to the other three methods of \eqnrtwo{eqn:pVal}{eqn:AIC} with and without the top 30\% fit cut is then added as a systematic uncertainty. An unrepresentative case would have the weight dominated by a single, outlying fit. If this is the case, that method is not used when determining the systematic uncertainty.

\subsection{Spectrum}

We are now in a position to present the spectrum of $J=1/2$ charmed baryons with positive and negative parity at the lowest temperature, in \Fig{fig:spectAll} and \Tab{tab:NT128-masses}. While exact agreement with the experimental results is not expected due to our heavier-than-physical light quarks and lack of continuum and infinite volume extrapolation, there is encouraging similarity in the observed masses in both the positive and negative parity channels.
Better agreement is observed for hadrons with more strange and charm quarks; this can be explained by the tuning of these heavier quarks to their physical masses~\cite{HadronSpectrum:2008xlg,HadronSpectrum:2012gic,Cheung:2016bym}.
We remind the reader that we do not attempt precision zero-temperature spectroscopy, see e.g.\ Refs.~\cite{HadronSpectrum:2012gic,Padmanath:2021lzf,Na:2006qz,Na:2007pv,Na:2008hz,Padmanath:2013zfa,Brown:2014ena,Perez-Rubio:2015zqb,Padmanath:2015jea,Chen:2017kxr,Mathur:2018rwu,Padmanath:2018zqw,Padmanath:2019ybu,Chiu:2020ppa,Bahtiyar:2022nqw} instead, but that our interest is in the response to heating up the system, to which we turn in the next section.

\section{Temperature Effects}
\label{sec:ratios}

To investigate the effect of increasing the temperature of the hadronic medium, we first attempt to assess this using only information contained in the correlators, i.e., without performing fits at finite temperature. We use two complementary approaches; first we perform an appropriate resummation of the correlator at the lowest temperature to account for the difference in temporal extent at finite temperature. This leads to the so-called \enquote{reconstructed} correlators \cite{Ding:2012sp}, extended to the fermionic case here. Subsequently, we consider a minimal model of the thermal correlator using information only from the well-defined ground state fits at the lowest temperature and take appropriate ratios, following Ref.~\cite{Aarts:2022krz}.
In both these approaches, the key idea is that we attempt to isolate the effect of changing the  temperature on the spectrum, while removing the effect of reducing the temporal extent as a trivial kinematic construction on the Euclidean lattice.

\subsection{Reconstructed Correlator}

The spectral relation for baryons reads, see e.g.~Eqs.~(2.16, 2.17, 2.27) of \Refl{Aarts:2017rrl},
\begin{align}
  G\rb{\tau; T} = \int_{-\infty}^{\infty}\frac{d\omega}{2\pi}\, K_F\rb{\tau, \omega; T} \rho\rb{\omega},
  \label{eqn:specRel}
\end{align}
where $K_F$ is the fermionic kernel
\begin{align}
  K_F\rb{\tau, \omega; T} = \frac{e^{-\omega\tau}}{1 + e^{-\omega / T}}.
\end{align}
Here $G\rb{\tau; T}$ and $\rho\rb{\omega}$ can have Dirac indices, or be projected with parity operators, see \Refl{Aarts:2017rrl}.
\par
This kernel can be compared with the building block for bosonic (mesonic) correlators
\begin{align}
  K_B\rb{\tau, \omega; T} = \frac{e^{-\omega\tau}}{1 - e^{-\omega / T}}.
\end{align}
Often this is multiplied with $\exp(\omega/2T)$ in the numerator and the denominator, to arrive at the usual $\sinh(\omega/2T)$ in the denominator, but this is not needed here.

To relate a correlator at a higher temperature $T$ to one at a lower temperature $T_0$ by resummation, we use a simple identity which follows from factorisation. We switch to lattice units such that $T=1/N_\tau$, $T_0=1/N_0$, and $N_0/N_\tau=m$ is an (odd) integer. In the bosonic case
\begin{align}
  1 - \expe{-\omega m N_{\tau}} = \rb{1 - \expe{-\omega N_{\tau}}}\,\rb{\sum_{n=0}^{m-1}\expe{-n\omega N_{\tau}}},
\end{align}
where $m$ is any integer. The fermion case is
\begin{align}
  1 + \expe{-\omega m N_{\tau}} = \rb{1 + \expe{-\omega N_{\tau}}}\,\rb{\sum_{n=0}^{m-1}\rb{-1}^{n}\,\expe{-n\omega N_{\tau}}}
\end{align}
for $m$ an odd integer. The bosonic kernel can hence be written as
\begin{align}
  K_B(\tau, & \, \omega;1/N_{\tau}) = \frac{\expe{-\omega \tau}}{1 - \expe{-\omega N_{\tau}}} = \sum_{n=0}^{m-1}\frac{\expe{-\omega\rb{\tau + n N_{\tau}}}}{1 - \expe{-\omega m N_{\tau}}} \nn \\
      &= \sum_{n=0}^{m-1}K_B\rb{\tau + n N_{\tau}, \omega; 1/(m N_{\tau})},
\end{align}
for integer $m$, and the fermionic kernel as
\begin{align}
\label{eq:reconF}
  K_F(\tau, & \,\omega;  1/N_{\tau}) = \frac{\expe{-\omega\tau}}{1 + \expe{-\omega N_{\tau}}} = \sum_{n=0}^{m-1}\frac{\rb{-1}^{n}\,\expe{-\omega\rb{\tau + n N_{\tau}}}}{1 + \expe{-\omega m N_{\tau}}} \quad \nn \\
  &= \sum_{n=0}^{m-1}\rb{-1}^{n}\,K_F\rb{\tau + n N_{\tau}, \omega; 1/(m N_{\tau})},
\end{align}
for integer odd $m$.

Inserting Eq.~(\ref{eq:reconF}) into the spectral relation  (\ref{eqn:specRel}) for a fermionic correlator at temperature $T=1/N_\tau$ relates this correlator to a resummation of one at a lower temperature $T_0=1/N_0 = 1/(m N_{\tau})$, assuming that the spectral content content is unchanged. Thus yields the reconstructed correlator for fermions, 
\begin{align}
  G_{\text{rec}}\rb{\tau; 1/N_\tau, 1/N_0} = \sum_{n=0}^{m-1}\rb{-1}^{n}\,G\rb{\tau + n N_{\tau}; 1/N_0}.
  \label{eqn:reconG}
\end{align}
Switching back to denoting the temperatures with $T$ and $T_0$, it allows us to compare the actual correlator at temperature $T$, $G(\tau; T)$, with the expected correlator if the spectral content is unchanged from a lower reference temperature $T_0$, namely $ G_{\text{rec}}\rb{\tau; T, T_0}$, via the ratio
\begin{align}
\label{eq:r-rec}
r_{\rm rec}(\tau; T, T_0) = G\rb{\tau; T} / G_{\rm rec}\rb{\tau; T, T_0}.
\end{align}
For bosons, this technique was used in e.g.\ Refs.~\cite{Ding:2012sp,Kelly:2018hsi}. For fermions, we are not aware of a previous application.

As $m=N_0/N_{\tau}=T/T_0$ must be an odd integer for fermions, the lattice sizes which can be used are limited in principle. This can be resolved by \enquote{padding} the correlator with the value at the minimum. No qualitative difference is found by padding with zero instead. We consider only adding points to extend the correlator as this can be done at all temperatures. In principle one could also consider removing points from the reference correlator at the lower temperature $T_0=1/N_0$. No substantial difference is observed in our tests.

\subsection{Model Correlator}

Alternatively, we may construct simple model correlators at a higher temperature $T$ by employing the ground state masses determined at the lowest temperature $T_0$ in the previous section, built on the assumption that spectral content has not changed and using the baryon spectral relation (\ref{eqn:specRel}) for the zero-momentum projected positive-parity correlator. We write for the spectral function 
\begin{align}
  \rho_+\rb{\omega} = 2\pi A_+\delta\rb{\omega - M^{+}_0} + 2\pi A_-\delta\rb{\omega + M^{-}_0},
\end{align}
where $M^\pm_{0}$ is the positive (negative) parity mass at temperature $T_0$ and $A_\pm$ are the corresponding amplitudes. The model correlator at temperature $T$ is then
\begin{align}
  G_{\rm model}\rb{\tau; T, T_0} &= A_+ K_F\rb{\tau, M^{+}_{0}} + A_-K_{F}\rb{\tau, -M^{-}_{0}} \nn \\
  &=\frac{A_+\expe{-M^{+}_{0}\tau}}{1 + \expe{-M^{+}_{0}/T}} + \frac{A_-\expe{M^{-}_{0}\tau}}{1 + \expe{M^{-}_{0}/T}}.
\end{align}
The main assumption here is that the width of the state extracted is negligible.
We can now take ratios of the actual and the model correlator, 
\begin{align}
  r\rb{\tau; T, T_0} = G\rb{\tau; T} / G_{\rm model}\rb{\tau; T, T_0},
  \label{eqn:SRatio}
\end{align}
as well as double ratios,
\begin{align}
  R\rb{\tau;T,T_0} & = \frac{r\rb{\tau; T, T_0}}{r\rb{\tau; T_0, T_0}} \nn \\
  &= \left.\frac{G\rb{\tau; T}}{G_{\rm model}\rb{\tau; T, T_0}}\right/\frac{G\rb{\tau; T_0}}{G_{\rm model}\rb{\tau; T_0, T_0}},
  \label{eqn:DRatio}
\end{align}
to potentially eliminate the effects of excited states and other features at early Euclidean times. This approach has been studied in some detail in Ref.~\cite{Aarts:2022krz} for the case of $D_{(s)}$ mesons. Note that in all ratios we may consider separately the positive parity correlator $G_+(\tau)$ and the negative parity correlator $G_-(\tau) = -G_+(1/T-\tau)$. Moreover,  we always divide the correlators by their value at $\tau=0$, to set a consistent scale.
For further analysis on the reconstructed and double ratios, see Ref.~\cite{bignell2023reconstructed}.

\begin{figure*}[!p]
  \centering
  \includegraphics[width=0.42\linewidth]{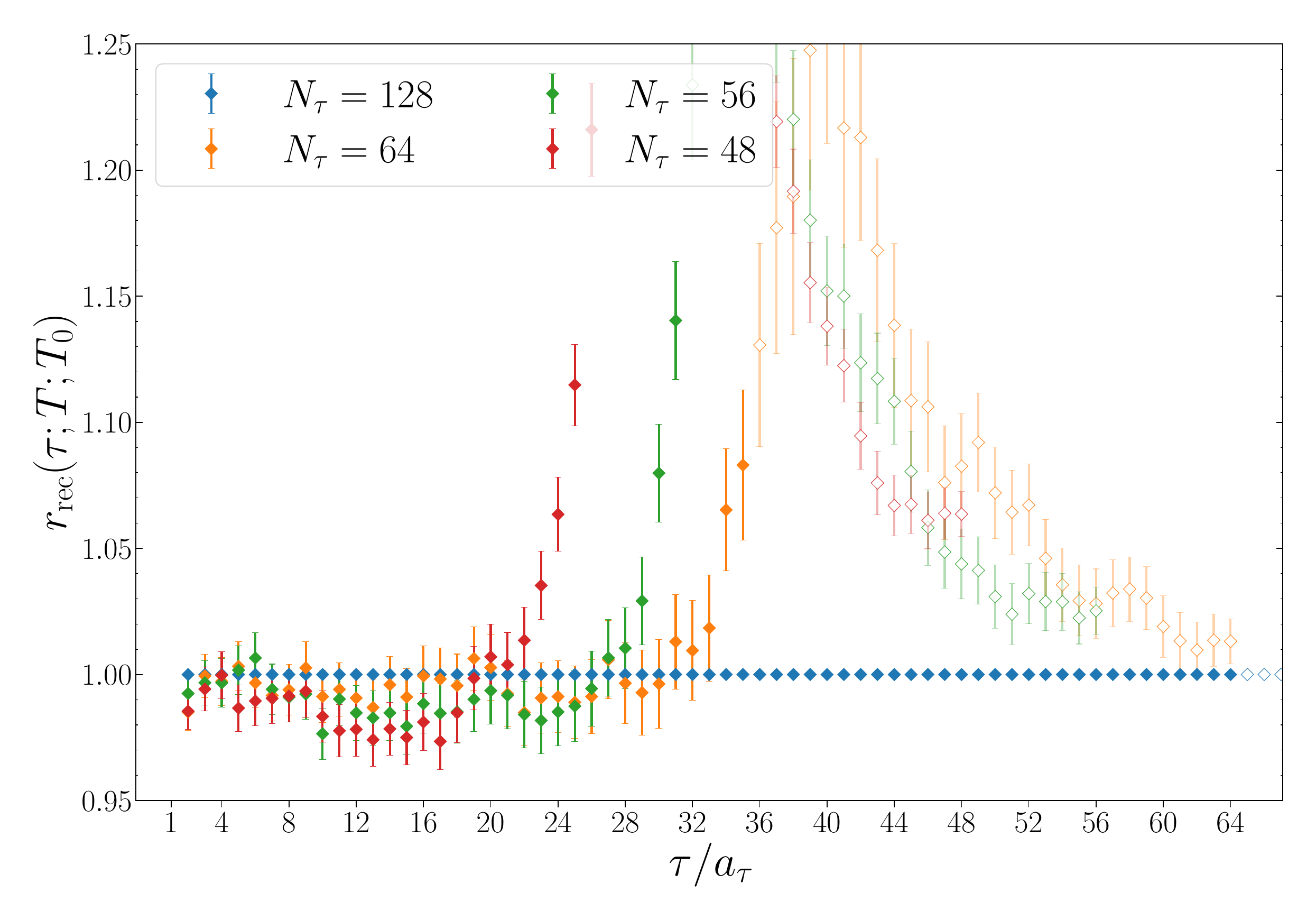}
  \includegraphics[width=0.42\linewidth]{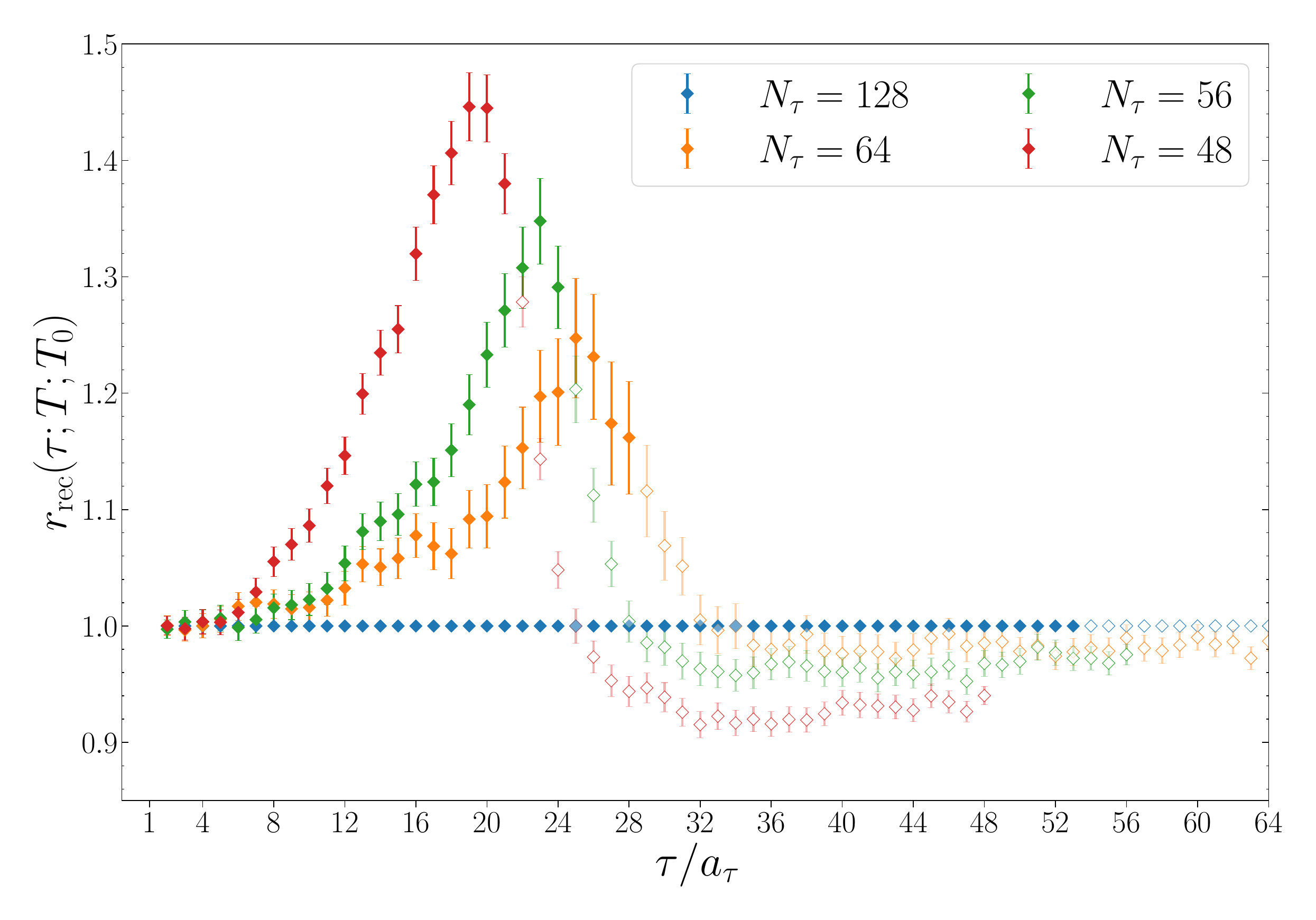}
  \\
  \vspace{-0.0053\textheight}
  \includegraphics[width=0.42\linewidth]{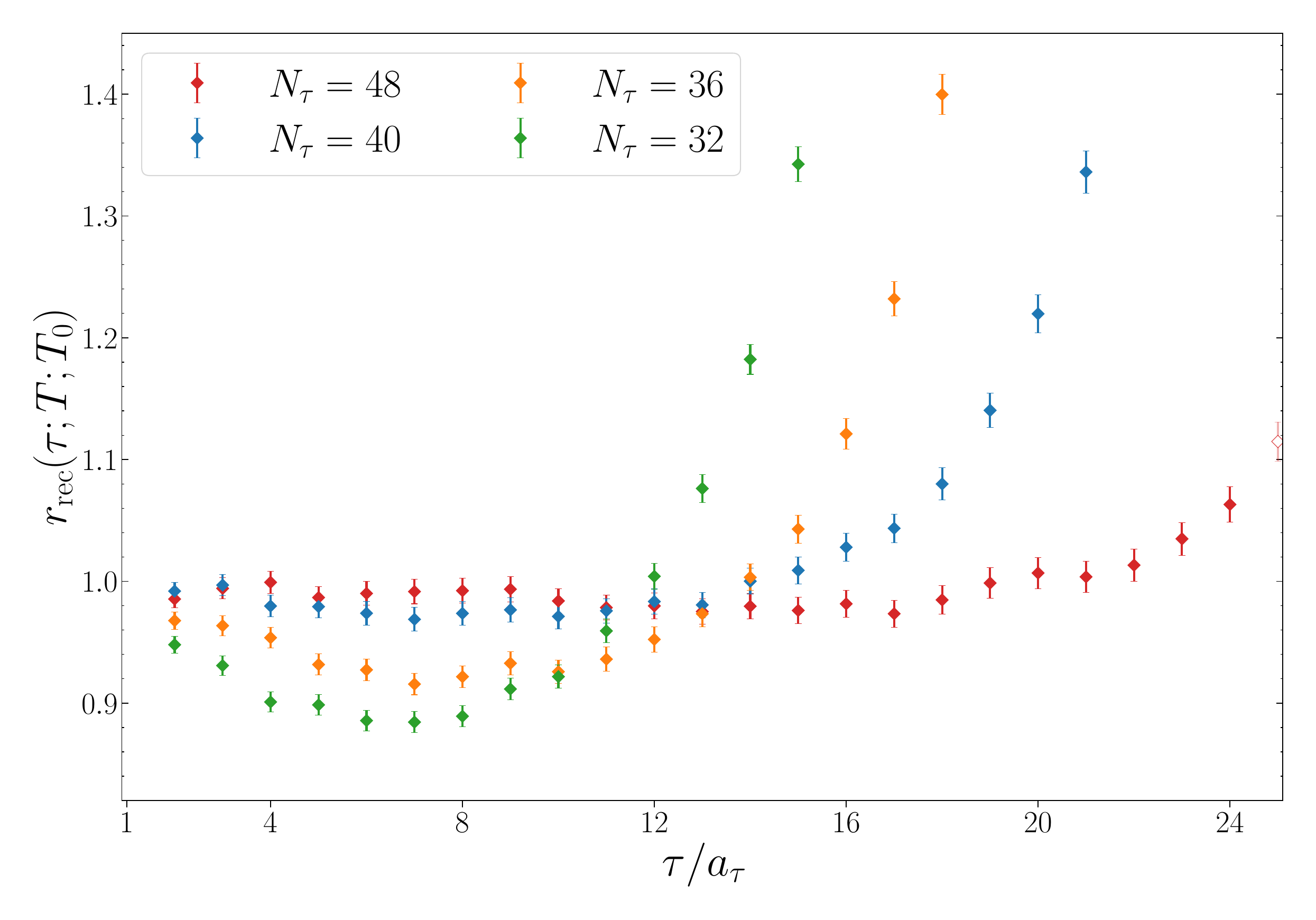}
  \includegraphics[width=0.42\linewidth]{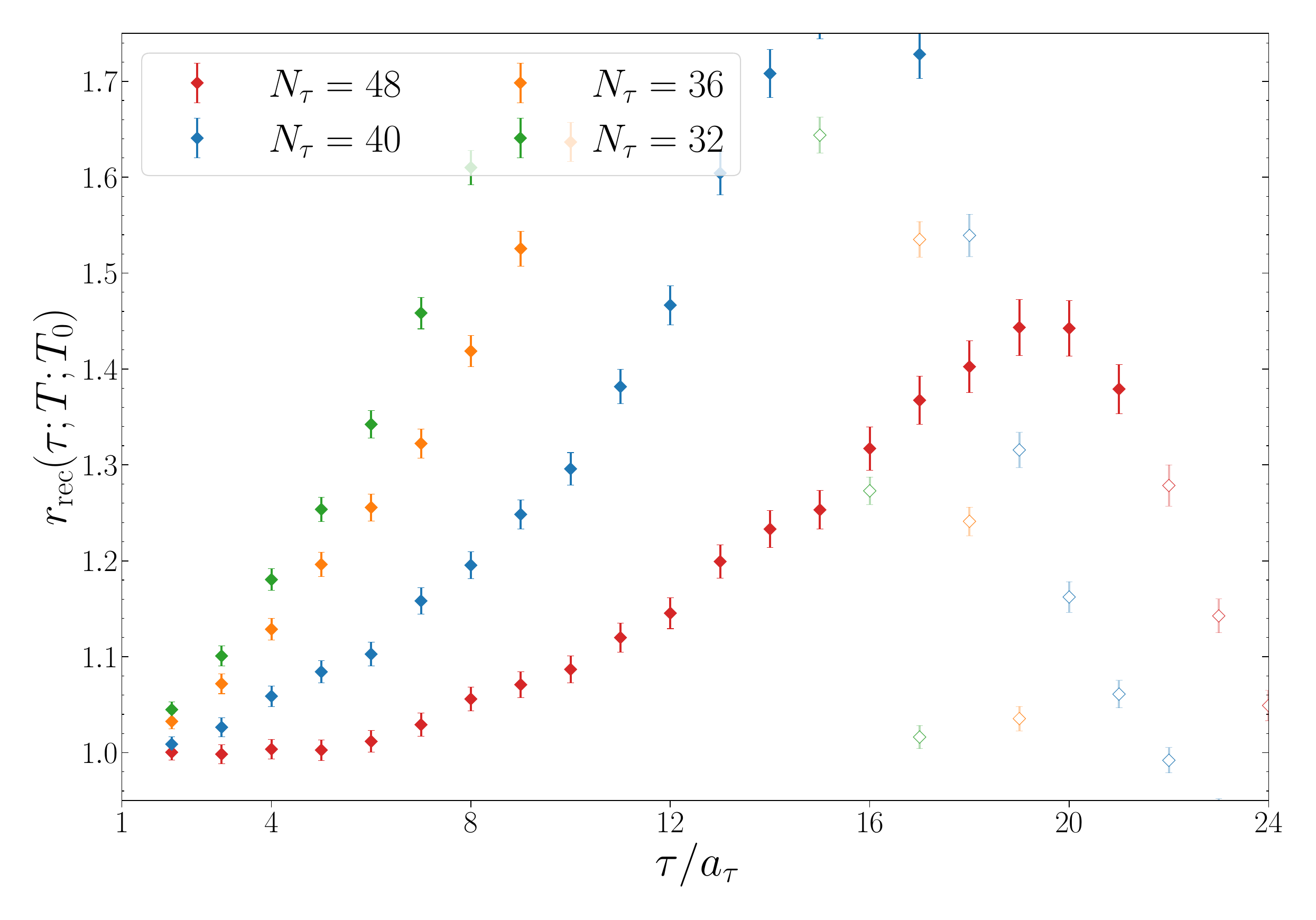}
  \caption{Ratio of correlator $G\rb{\tau; T}$ and reconstructed correlator $G_{\text{rec}}\rb{\tau; T, T_0}$ as described in Eq.~\ref{eq:r-rec} in the $\Sigma_{c}\rb{udc}$ channel, using as input the correlator at the lowest temperature $T_0=47$ MeV ($N_0=128$), for positive parity (left) and negative parity (right). The top row shows the four lowest temperatures, $T=47, 95, 109, 127$ MeV; the bottom row the higher ones, $T=127, 152, 169, 190$ MeV.
  Points past the minimum of the lattice correlator $G\rb{\tau; T}$ are faded and have open symbols.
  \label{fig:sigma12_3fl_udc_single_recon}
  }

\vspace{0.5cm}

  \centering
  \includegraphics[width=0.42\linewidth]{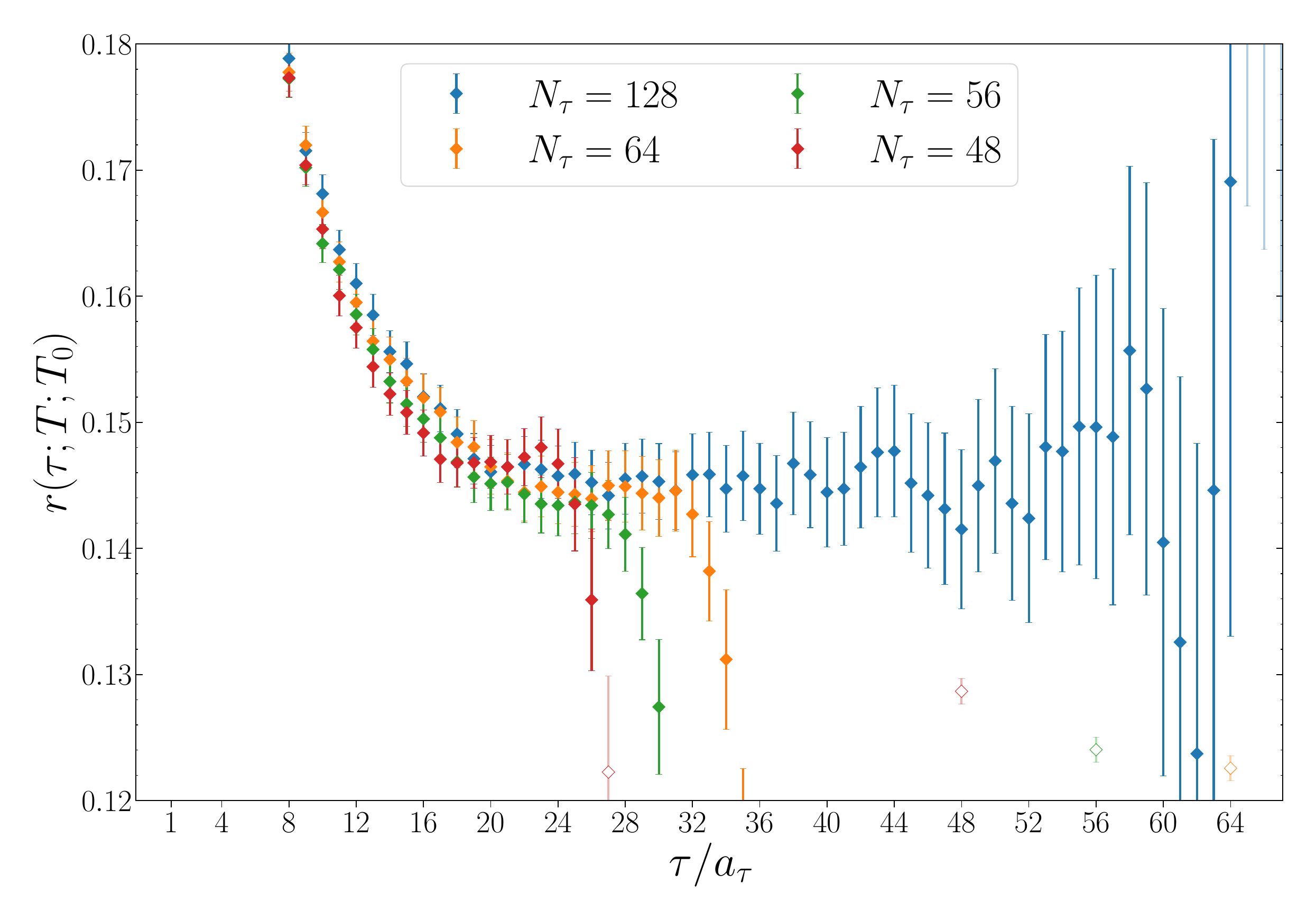}
  \includegraphics[width=0.42\linewidth]{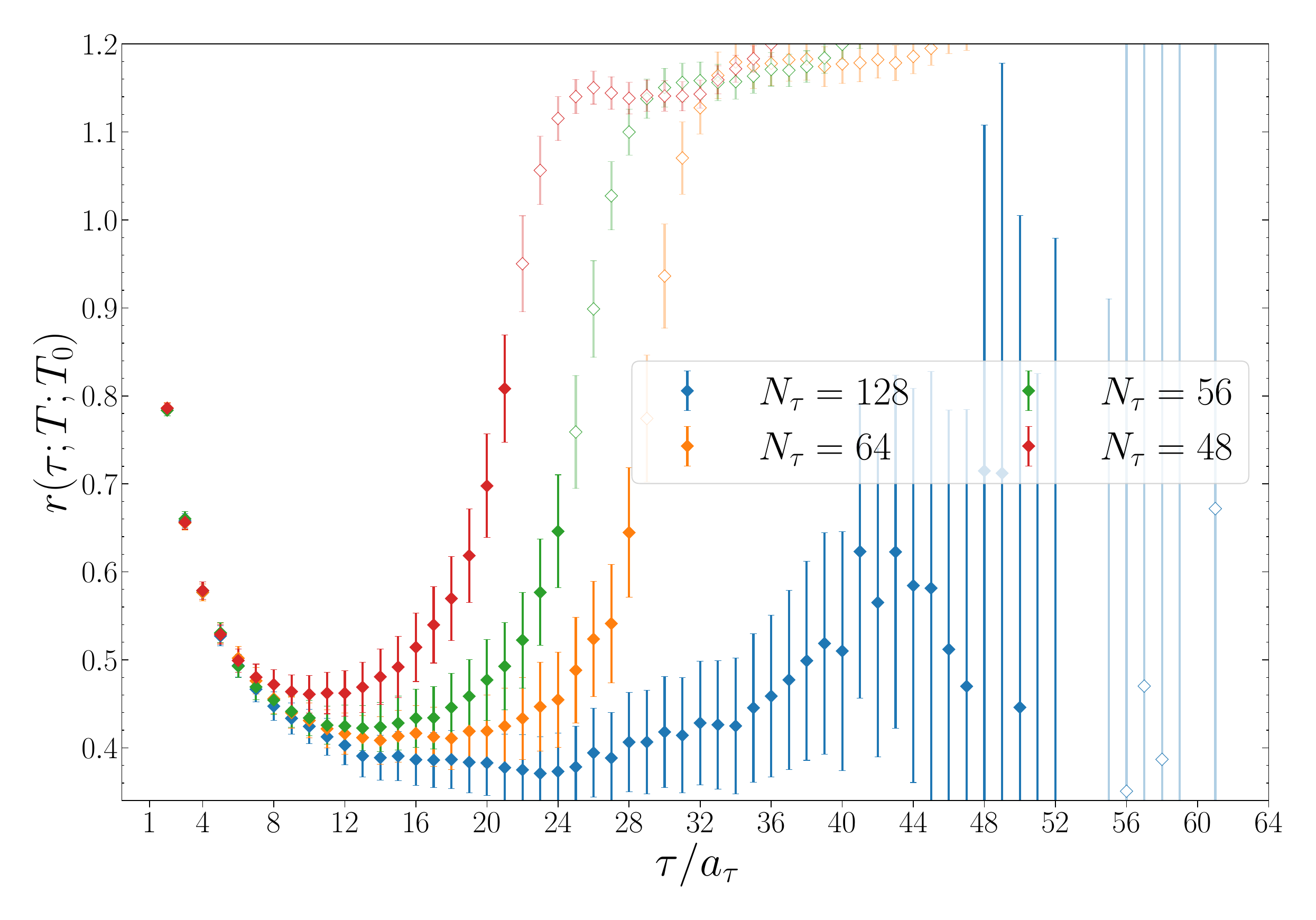}
  \\
  \vspace{-0.0053\textheight}
  \includegraphics[width=0.42\linewidth]{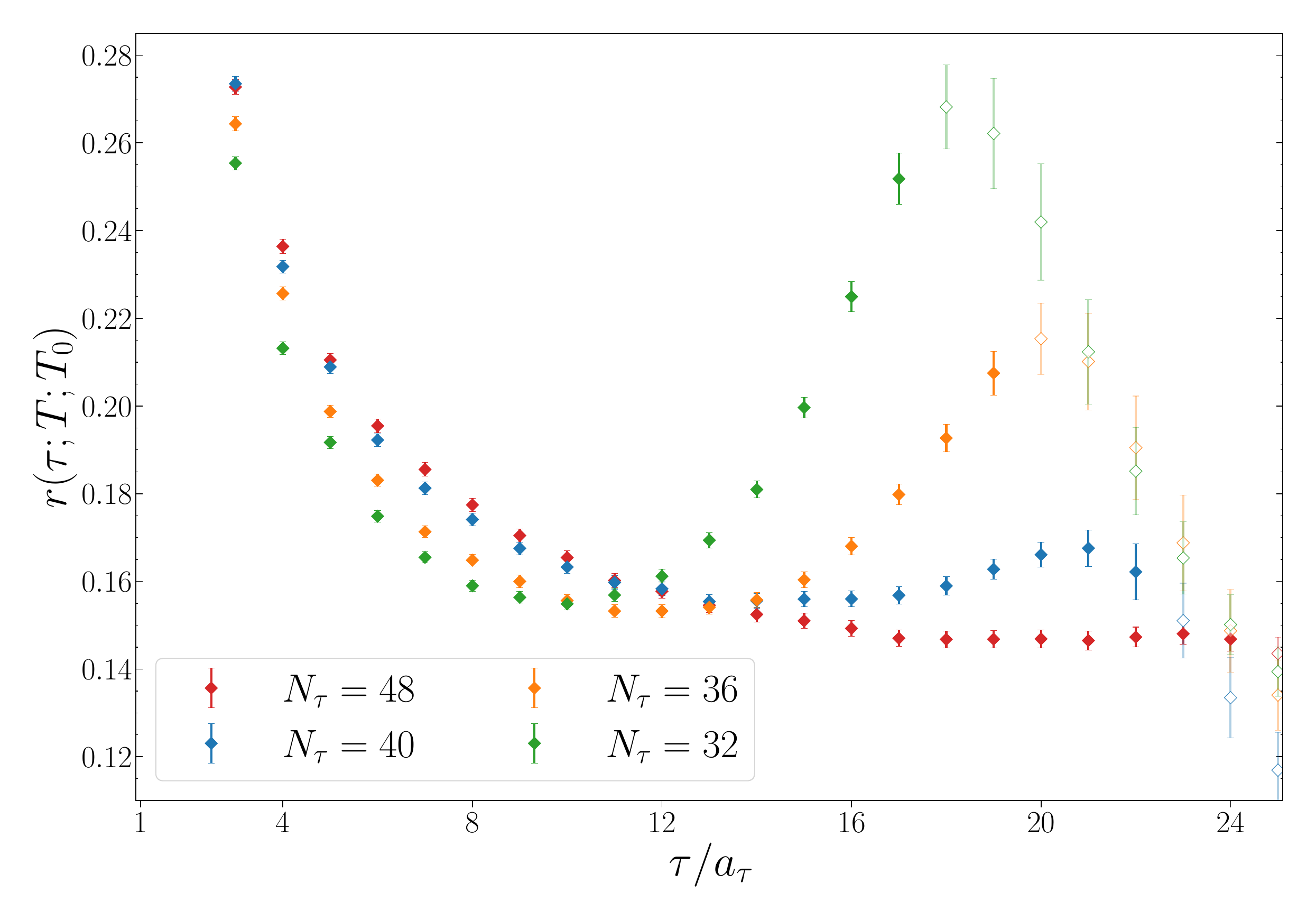}
  \includegraphics[width=0.42\linewidth]{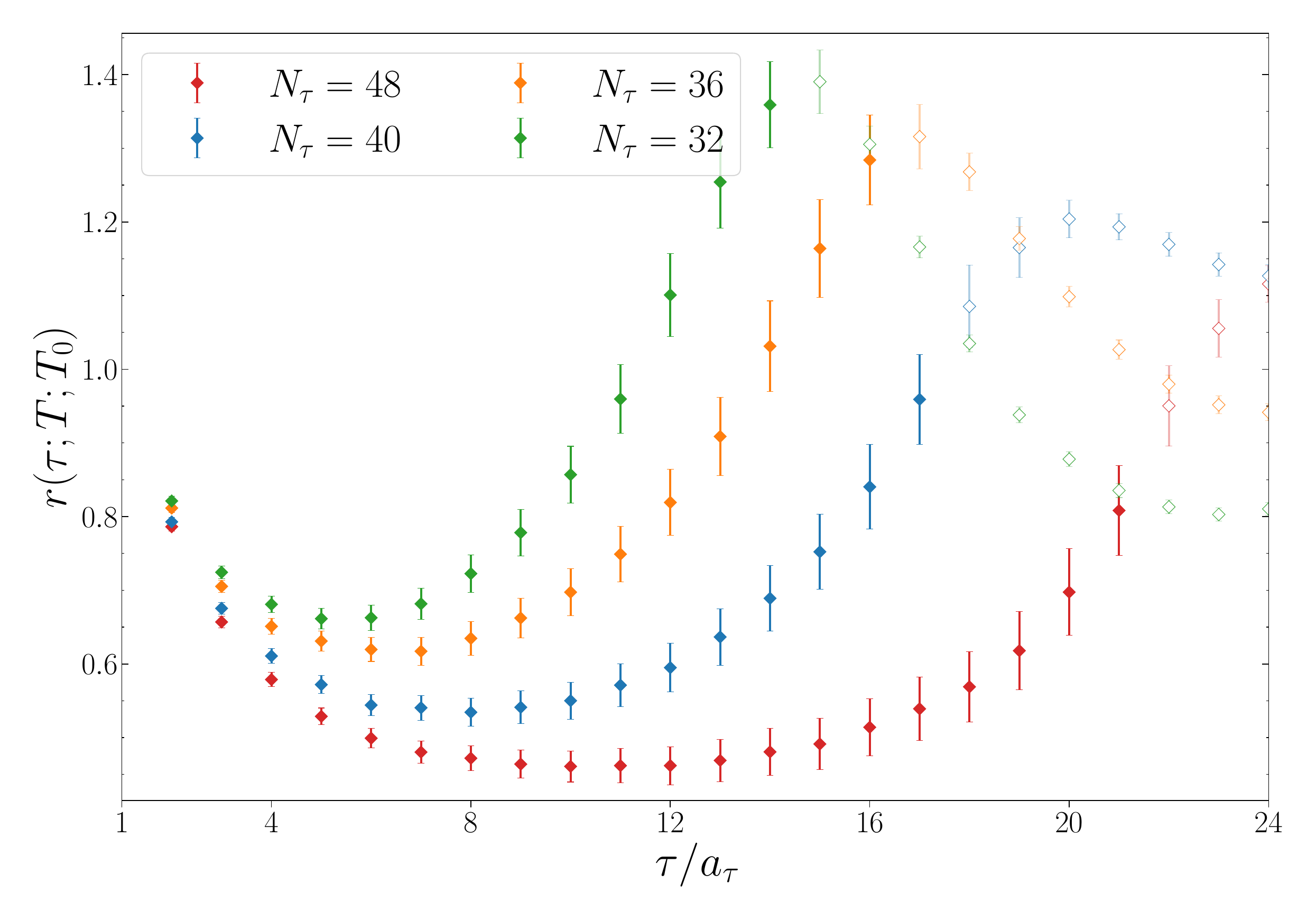}
  \caption{As in Fig.~\ref{fig:sigma12_3fl_udc_single_recon}, for the ratio of the correlator $G\rb{\tau; T}$ and the model correlator $G_{\rm model}\rb{\tau; T, T_0}$, see \eqnr{eqn:SRatio}.
  }
  \label{fig:sigma12_3fl_udc_single}
\end{figure*}

\subsection{Results}

We now present some results for ratios of correlators, using both the reconstructed correlator and the model correlator in the denominator. We focus here on the $\Sigma_{c}\rb{udc}$ channel; Appendix \ref{sec:app} contains additional results in the $\Xi_{cc}\rb{ccu}$ and $\Omega_{cc}\rb{ccs}$ channels. These are representative; we have of course analysed all channels.

We start with the ratio using the reconstructed correlators, see Eq.~(\ref{eq:r-rec}).
In \Fig{fig:sigma12_3fl_udc_single_recon} we show the result in the $\Sigma_{c}\rb{udc}$ channel, for positive parity (left) and negative parity (right), at low temperatures (top) and at higher temperatures (bottom). Note that $T=127$ MeV is shown in both sets of figures. In the positive parity case, we note that the ratio is close to 1 at the lower temperatures; the bending upwards around $\tau/a_\tau = N_\tau/2$ is expected due to the appearance of states with opposite parity. At the higher temperatures, the deviation becomes more significant around $T\sim \Tpc$ and is of the order of 10\% at $T=190$ MeV. Note the different vertical scales.
We conclude that in the positive parity sector thermal effects are at the percent level up to $T\sim \Tpc$. In the negative parity case on the other hand, there are strong temperature effects at all temperatures, even deep in the hadronic phase.

The reconstructed correlator gives insight into possible changes into the underlying spectral functions as the temperature increases. A closer look at the fate of the ground state can be obtained by taking the ratio with the model correlator, see Eq.~(\ref{eqn:SRatio}). This analysis relies on the assumption that at the lowest temperature the width of the state extracted is negligible and that it is possible to distinguish the ground state from the excited states. As the temperature increases the temporal extent of the lattice decreases and so the ground state may not be distinguishable from the excited states, even if the spectral information is unchanged. 
If the ground states are precisely determined and indeed independent of temperature, the ratio Eq.~(\ref{eqn:SRatio}) will be independent of $\tau$ for large enough Euclidean time, understood as the region where the ground state dominates, but where the \enquote{bending} of the correlator near $N_\tau/2$ is not yet encountered. As the model contains only the ground state and the correlator $G\rb{\tau; T}$ has excited states present, this ratio is not expected to be completely independent of $\tau$. 

This is demonstrated in Figure~\ref{fig:sigma12_3fl_udc_single}, again in the $\Sigma_{c}\rb{udc}$ channel. On the positive parity side, we do indeed observe a horizontal plateau for $T \lesssim 127 $ MeV, indicating that the ground state is well described by a single exponential. 
The negative parity sector is less clear due to the skewness of the correlation function causing fewer time slices to be available before the mixing of forward and backward states becomes evident. This could be improved by using methods to reduce the amount of excited state contamination in the correlator. 
The single ratio is similar to an effective mass in that it shows when excited state effects cannot be ignored.
The presence of plateaus in the single ratio is indicative that the narrow peak ansatz for the spectral function is suitable for exponential fits as in \eqnr{eqn:expFit}. At our lowest temperature, $T=47$ MeV, excellent plateau behaviour is seen for all channels investigated, providing further support for the results for the ground states masses presented in \Tab{tab:NT128-masses}.

\begin{figure*}[t]
  \centering
  \includegraphics[width=0.425\linewidth]{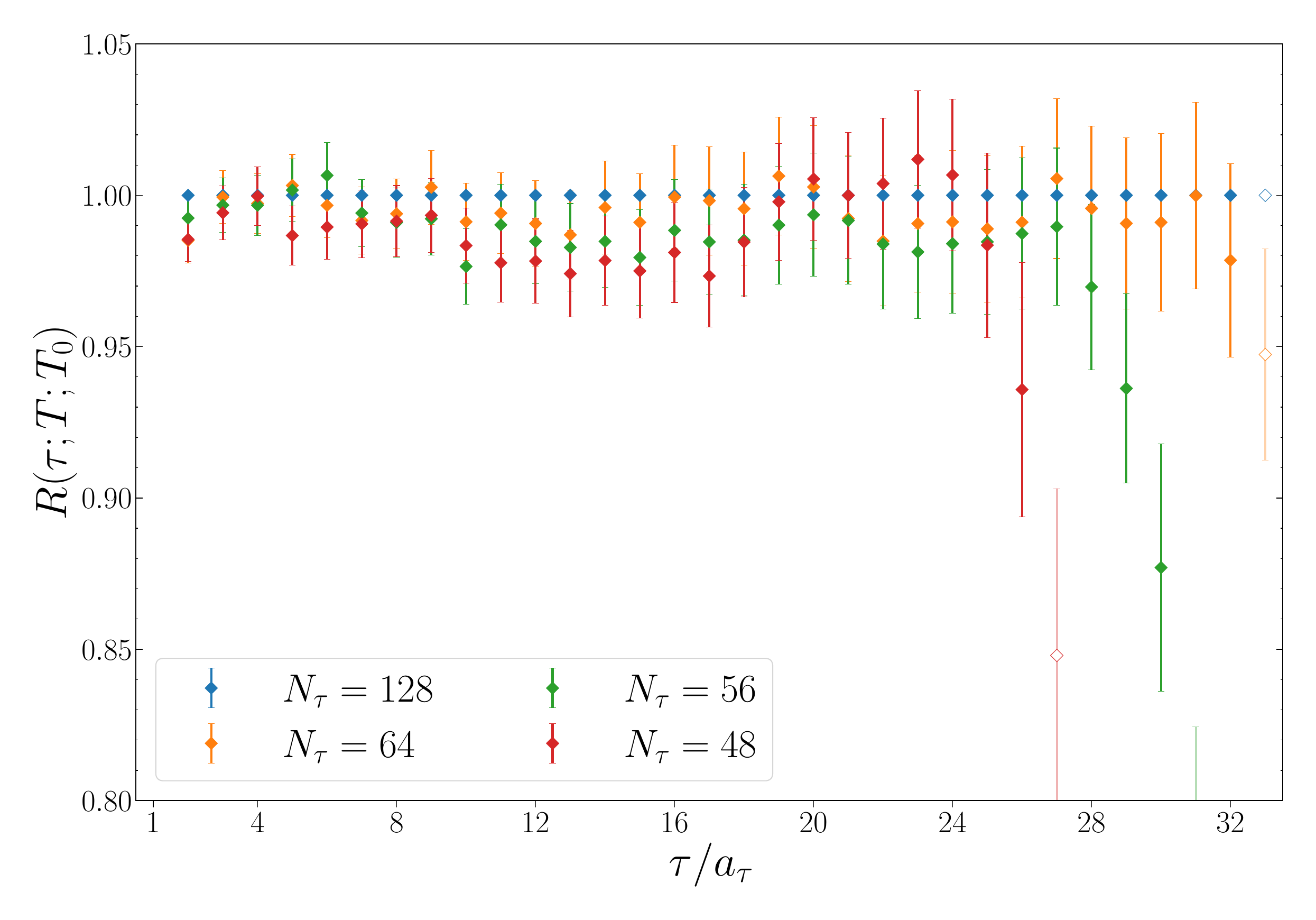}
  \includegraphics[width=0.42\linewidth]{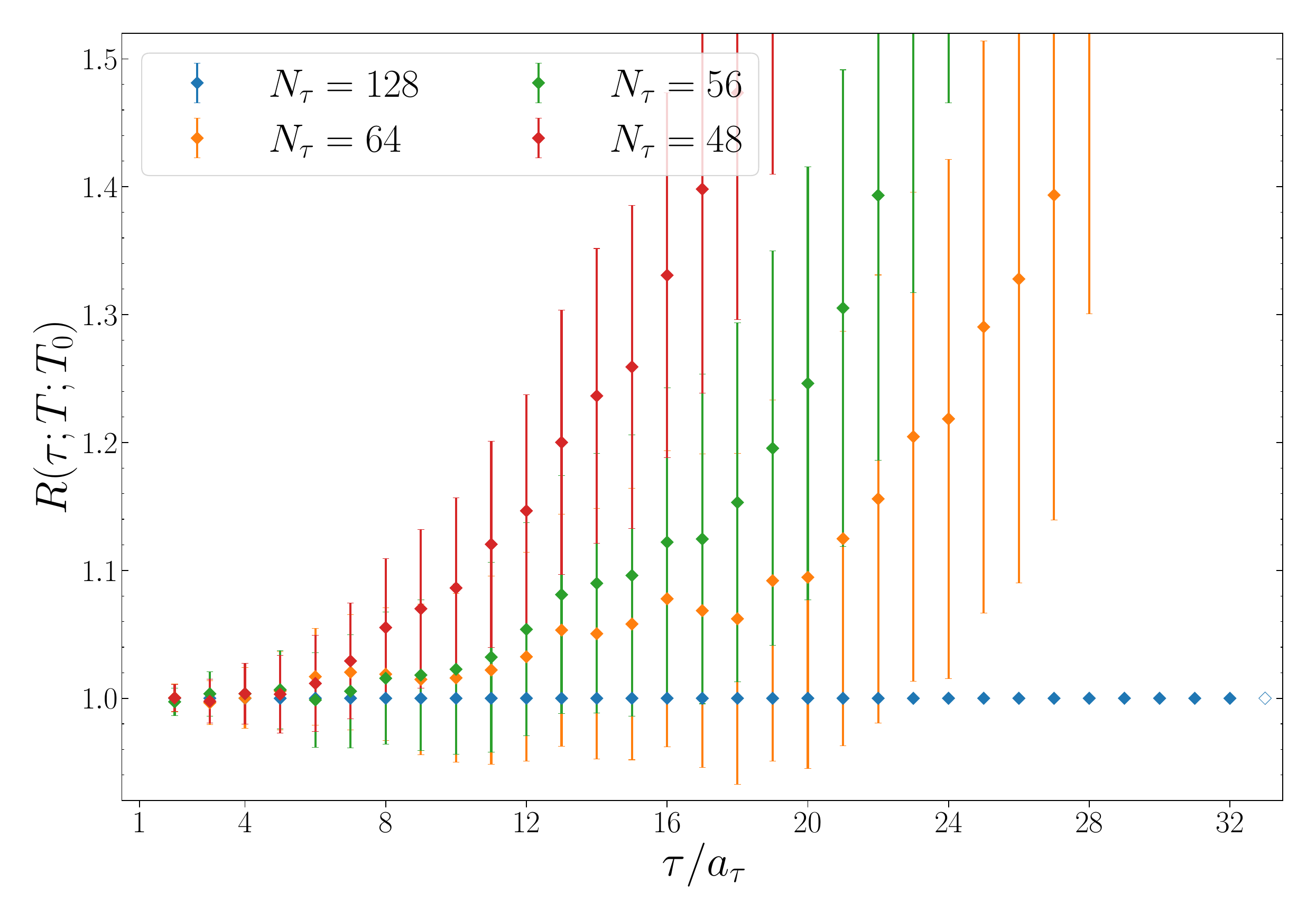}
  \\
  \vspace{-0.0053\textheight}
  \includegraphics[width=0.42\linewidth]{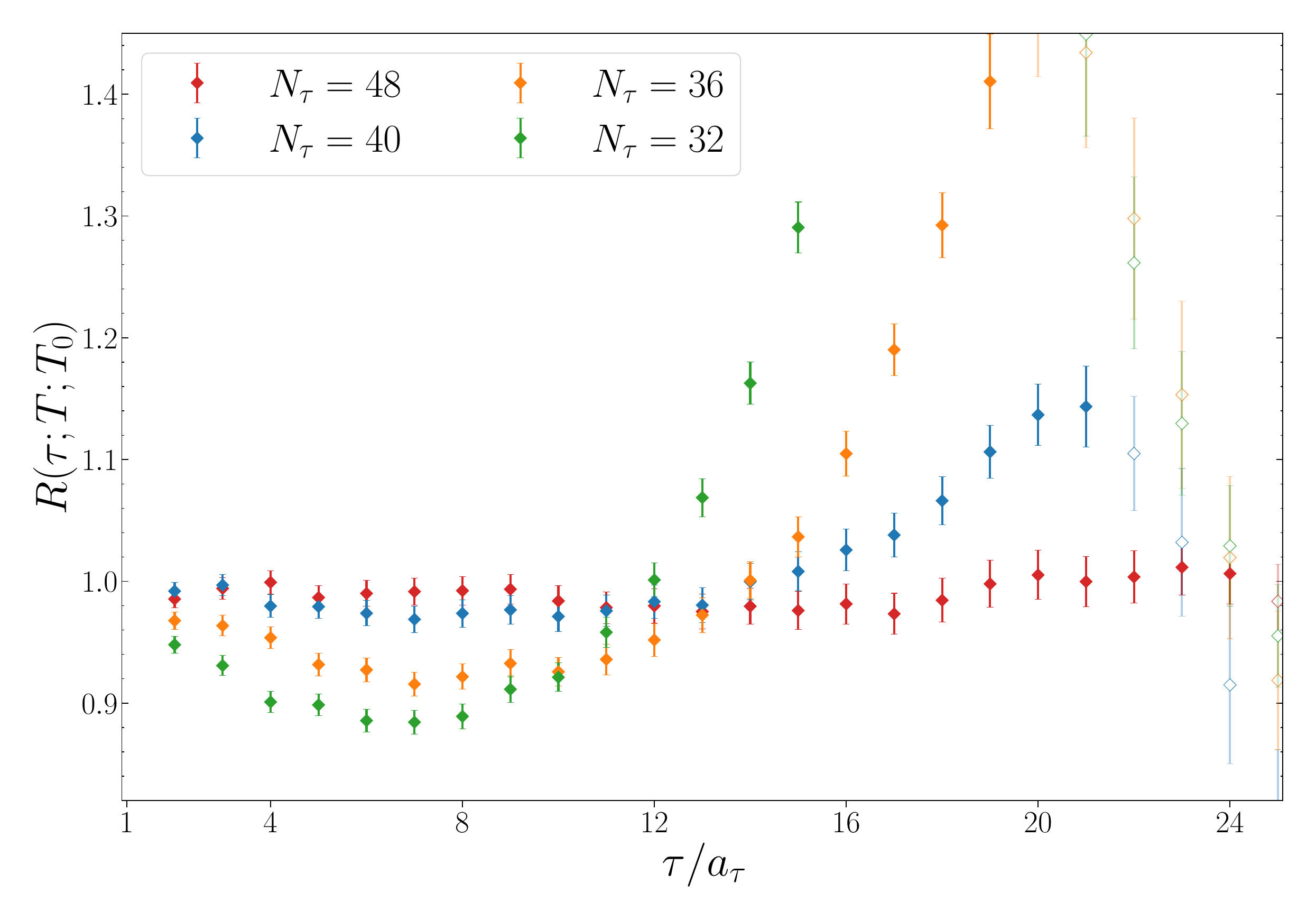}
  \includegraphics[width=0.42\linewidth]{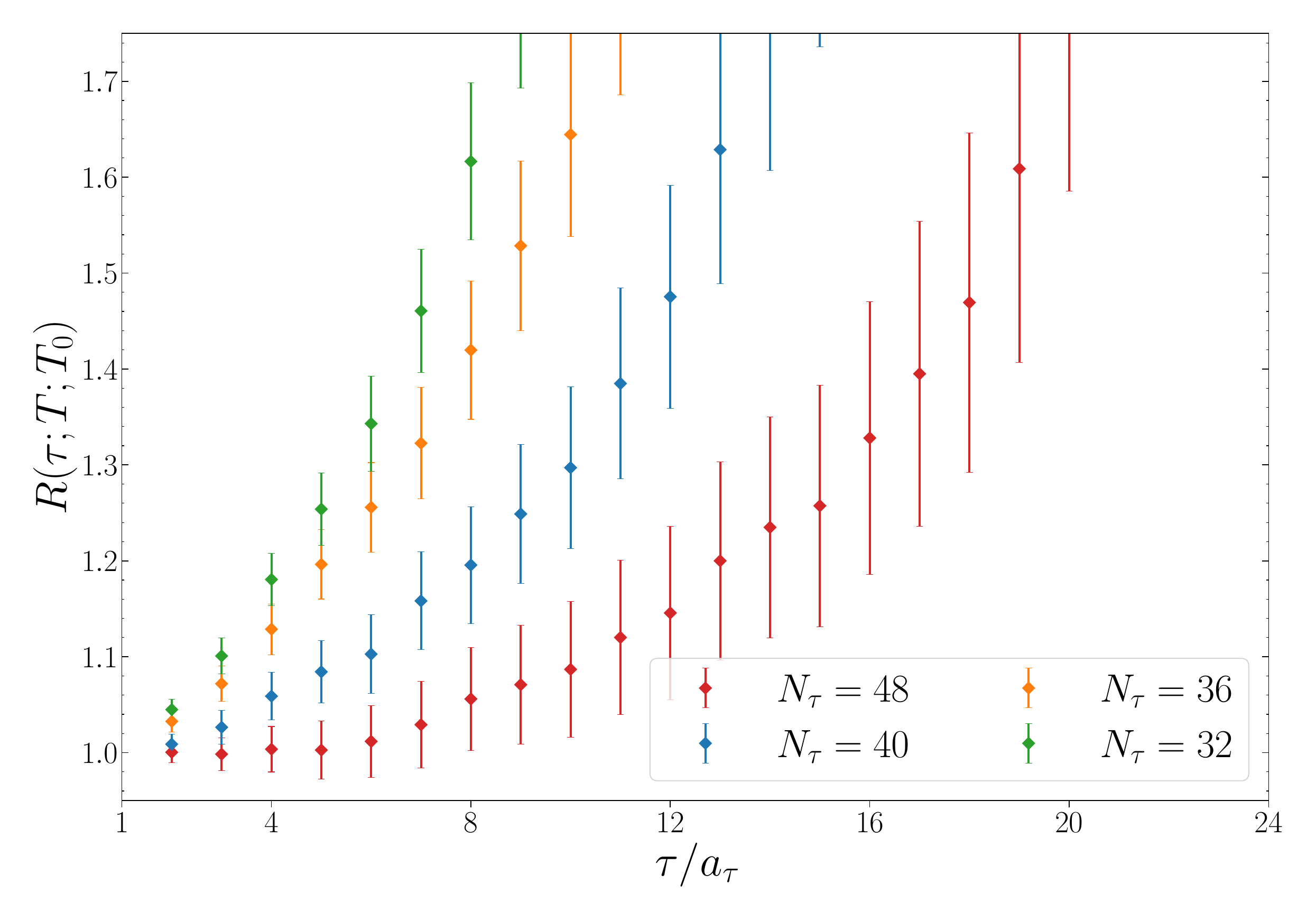}
  \caption{As in Fig.~\ref{fig:sigma12_3fl_udc_single_recon}, for the double ratio of $G\rb{\tau; T}/G_{\rm model}\rb{\tau; T, T_0}$ and $G\rb{\tau; T_0}/G_{\rm model}\rb{\tau; T_0, T_0}$, see \eqnr{eqn:DRatio}.
  }
  \label{fig:sigma12_3fl_udc_double}
\end{figure*}

From the plots with single ratios, it is clear that also the structure at earlier Euclidean times is approximately temperature independent, in particular at lower temperatures and in the positive parity sector, see Figure~\ref{fig:sigma12_3fl_udc_single} (top left).  
We can factor out the common behaviour by  considering the double ratio of \eqnr{eqn:DRatio}, involving temperatures  $T$ and $T_0$. 
The results are shown in Fig.~\ref{fig:sigma12_3fl_udc_double}.
We note that the magnitude of deviation from 1 is comparable between the double ratio and the ratio with the reconstructed correlator. Quantitatively, they show different behaviour: as mentioned, the double ratio only uses the information on the ground state at $T=T_0$, assuming that it is well described by a narrow peak, whereas the reconstructed correlator uses the full spectral function at $T_0$.

To determine where exponential fits can be reasonably performed to extract the ground state mass in the thermal case, we will require plateau-like behaviour with the double ratio not exceeding a difference of $\sim 10\%$ from one. This is illustrated in Fig.~\ref{fig:sigma12_3fl_udc_double}, where evidence can be seen of a change in behaviour around the pseudocritical temperature in the positive parity channel.
The double ratio is close to one until $T=169 \text{ MeV} \sim \Tpc$. The negative parity sector is much noisier --- note the different scale --- and deviates from one at much lower temperatures, removing the justification of using exponential fits for the negative parity state above $T = 127$ MeV in this channel.
When the negative parity sector has clear evidence of a change in spectral content, we do not report a negative parity mass obtained by fitting to an exponential Ansatz, as it is not justified to do so.
As mentioned, Appendix \ref{sec:app} contains additional results in the $\Xi_{cc}\rb{ccu}$ and $\Omega_{cc}\rb{ccs}$ channels.

\section{Masses at non-zero temperature}
\label{sec:masses}

\begin{figure}[t]
  \includegraphics[width=\columnwidth]{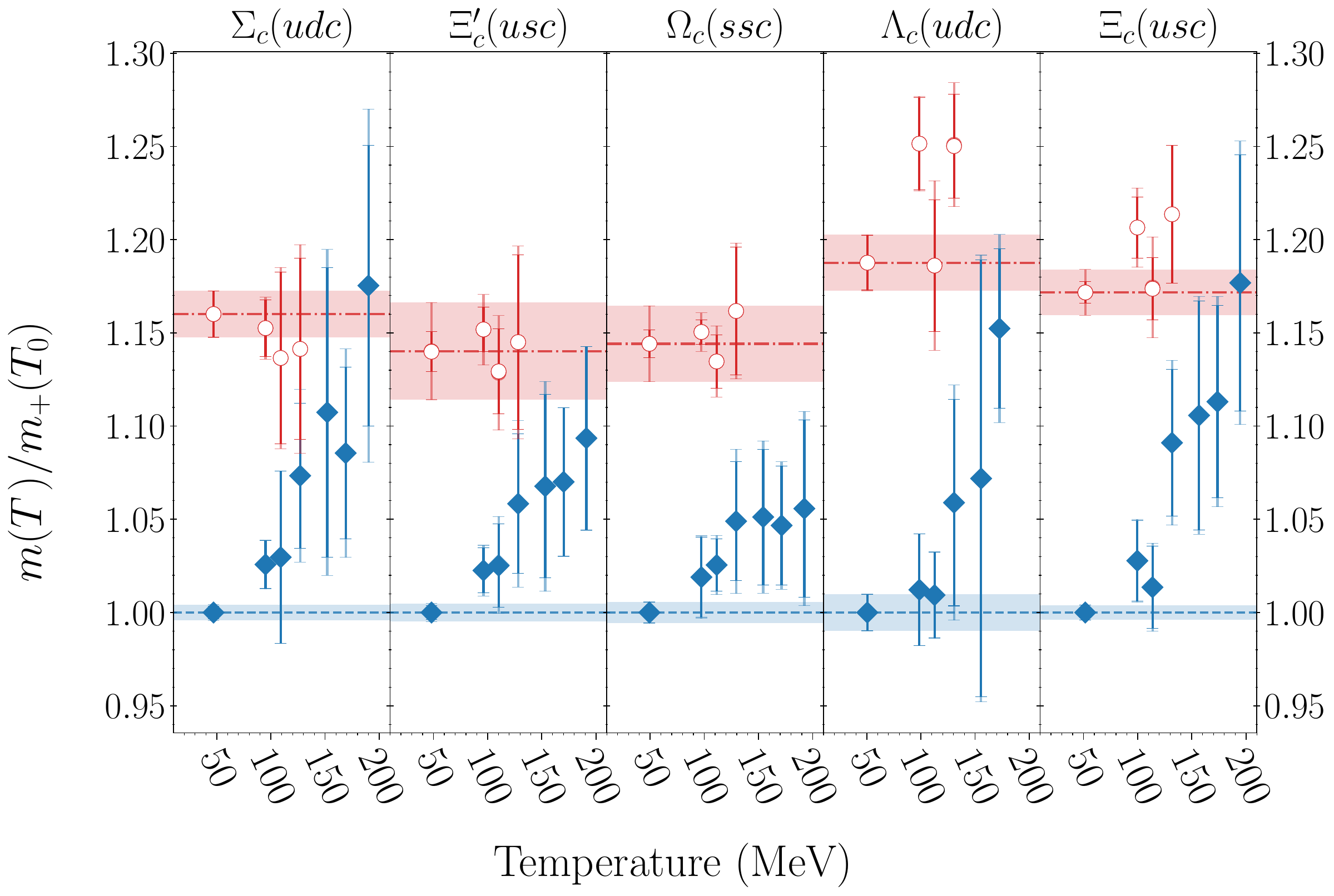}
  \caption{\label{fig:spectJ1_2P}Ground state masses of singly charmed spin $1/2$ baryons, normalised with the positive parity ground state mass at the lowest temperature, as a function of temperature. Filled (open) symbols are used for positive (negative) parity states. 
  The inner error bar represents the statistical uncertainty and the outer incorporates the systematic from the choice of averaging method. Horizontal dashed lines show the result from the lowest temperature. The uncertainty of the lowest temperature reflects the relative uncertainty of the lowest temperature mass.
  Masses are shown only when the analysis of \Sec{sec:ratios} suggests the mass can be extracted using an exponential Ansatz. 
  }
\end{figure}

We now extend the analysis detailed in \Sec{sec:Mass} to non-zero temperature. We use the analysis and methods of the previous sections as a guide to when the spectral content has changed such that the exponential Ansatz of \eqnr{eqn:expFit} is no longer appropriate. Consequently we show mass results for only a subset of the temperatures available to us. The fitting method used is identical to those in \Sec{sec:Mass}; the ratio analysis of the previous section provides guidance up to which temperature such an analysis can be applied.

The singly charmed baryon spectrum is presented in \Fig{fig:spectJ1_2P}. All masses have been normalised with the positive parity ground state mass at the lowest temperature, $m_+(T_0)$, in the corresponding channel.
While the uncertainties increase as temperature does --- reflective of fewer temporal points and a potential shift away from a zero-width state --- the pattern suggests that in the positive parity sector the masses increase with temperature and that this starts relatively deep in the confining phase. On the negative parity side, acceptable results are obtained, even though the negative parity signal is shorter lived than the positive parity, as illustrated in e.g.\ \Fig{fig:effMass}. Here the increase in uncertainties is such that it is difficult to determine if masses of negative parity states are changing in a systematic way. We note that the ratio analysis of \Sec{sec:ratios} did indeed suggest that the negative parity sector is more strongly affected by temperature effects and as such 
we do not report masses at as high a temperature for the negative parity states as compared to the positive parity ones.

\begin{figure}[t]
  \includegraphics[width=\columnwidth]{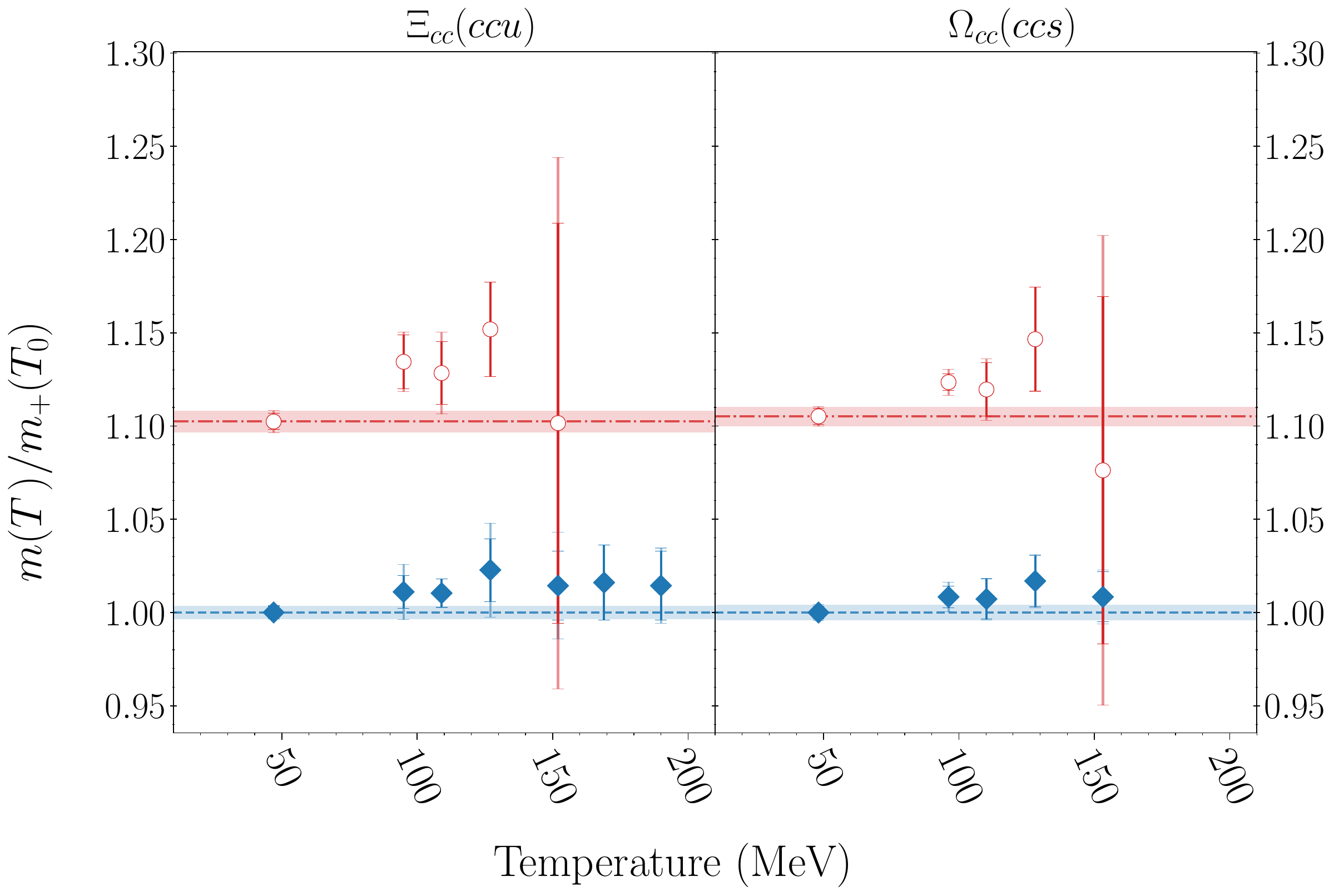}
  \caption{\label{fig:spectJ1_2M}As in Fig.~\ref{fig:spectJ1_2P}, for doubly charmed spin $1/2$ baryons.}
\end{figure}

Masses of the doubly charmed baryons $\Xi_{cc}\rb{ccu}$ and $\Omega_{cc}\rb{ccs}$ are shown in \Fig{fig:spectJ1_2M}. Temperature effects are much less evident here, in both parity sectors, with the positive parity $\Xi_{cc}\rb{ccu}$ mass remaining approximately constant well into the deconfining phase, up to $T\sim190$ MeV. To provide support for this finding we show the single and double ratios for this baryon in \Figtwo{fig:doublet_2fl_ccu_single}{fig:doublet_2fl_ccu_double} in Appendix \ref{sec:app}.
The positive parity double ratio in \Fig{fig:doublet_2fl_ccu_double} is particularly striking, with a value close to one at all temperatures shown here, indicating the apparent absence of thermal effects in this channel.
It would be interesting to further investigate the difference in thermal effects for doubly charmed baryons compared to singly charmed baryons using effective potential models, similar to the well-studied meson (quarkonium) cases; preliminary work in this direction has been carried out~\cite{Stasiakthesis}. 

While there is a possibility that new continuum states emerge in each of the channels considered at or near each respective ground state and conspire to keep the (double) ratio mostly unchanged, we consider this unlikely to occur in each channel. The simpler solution is a modification of the ground state masses. At a single volume, or without spectral function reconstruction methods, it is not possible for us to eliminate the possibility of new continuum states.

\section{Parity doubling}
\label{sec:par}

The restoration of chiral symmetry is a signature for the formation of a QGP~\cite{Borsanyi:2010cj,Bazavov:2011nk,Cheng:2010fe,Banerjee:2011yd}. We investigate chiral symmetry restoration via the parity doubling phenomenon~\cite{Aarts:2017rrl}, in which positive and negative parity correlation functions become degenerate when chiral symmetry is unbroken. While correlators for light and strange baryons exhibit a clear parity doubling signal~\cite{Aarts:2017rrl,Aarts:2018glk}, this is not expected here as the charm quark mass strongly and explicitly breaks chiral symmetry.
Nevertheless, the light quarks contained within the charmed baryons are susceptible to chiral symmetry, resulting in a restoration signal as we now demonstrate.

We define the baryon $R$ parameter~\cite{Datta:2012fz,Aarts:2015mma,Aarts:2017rrl,Aarts:2018glk,Aarts:2020vyb} from the summed ratio of positive and negative correlators,
\begin{align}
  \mcR\rb{\tau} &= \frac{G_+\rb{\tau} -G_+\rb{1/T-\tau}}{G_+\rb{\tau} + G_+\rb{1/T-\tau}}, \label{eqn:mcRratio} \\
  R\rb{n_0} &= \frac{\sum_{n=n_0}^{\frac{1}{2}N_\tau -1}\, \mcR\rb{\tau_{n}}/\sigma_{\mcR}^2\rb{\tau_{n}}}{\sum_{n=n_0}^{\frac{1}{2}N_\tau -1}\,1/\sigma_{\mcR}^2\rb{\tau_{n}}}.
  \label{eqn:Rratio}
\end{align}
Here $G_+\rb{\tau}$ is the positive parity correlator, $G_+\rb{1/T - \tau} = -G_-\rb{\tau}$ is the negative parity correlator --- see Eq.~(\ref{eq:Gpm}), $\sigma_{\mcR}\rb{\tau_{n}}$ denotes the statistical error for $\mcR\rb{\tau_{n}}$ and we consider a sum over time slices $\tau_{n}/a_\tau\in \sq{n_0, N_{\tau} / 2 - 1}$ at all temperatures. We choose $n_0=4$ to suppress lattice and excited state artefacts at small $\tau_{n}$~\cite{Aarts:2015mma}. Shifting $n_0$ slightly does not have a qualitative effect on the results and Gaussian smearing at the source and sink further suppresses the effect of excited states.
Lattice artefacts are due to e.g.\ the chiral symmetry violating Wilson term at larger energy scales.

As suggested by \eqnr{eqn:mcRratio}, the $R$ parameter describes the difference between the positive and negative parity correlators. When chiral symmetry is unbroken, the correlators are degenerate and $R=0$. Broken chiral symmetry produces $R\ne0$. In the limit where the correlators are dominated by their respective ground states and the positive parity mass is much smaller than the negative parity mass, $R\approx 1$. We hence expect that $R$ is close to one in the hadronic phase. Since the charm mass breaks chiral symmetry at all temperatures considered, we do not expect $R$ to go to zero.

\begin{figure}[tb]
  \includegraphics[width=\columnwidth]{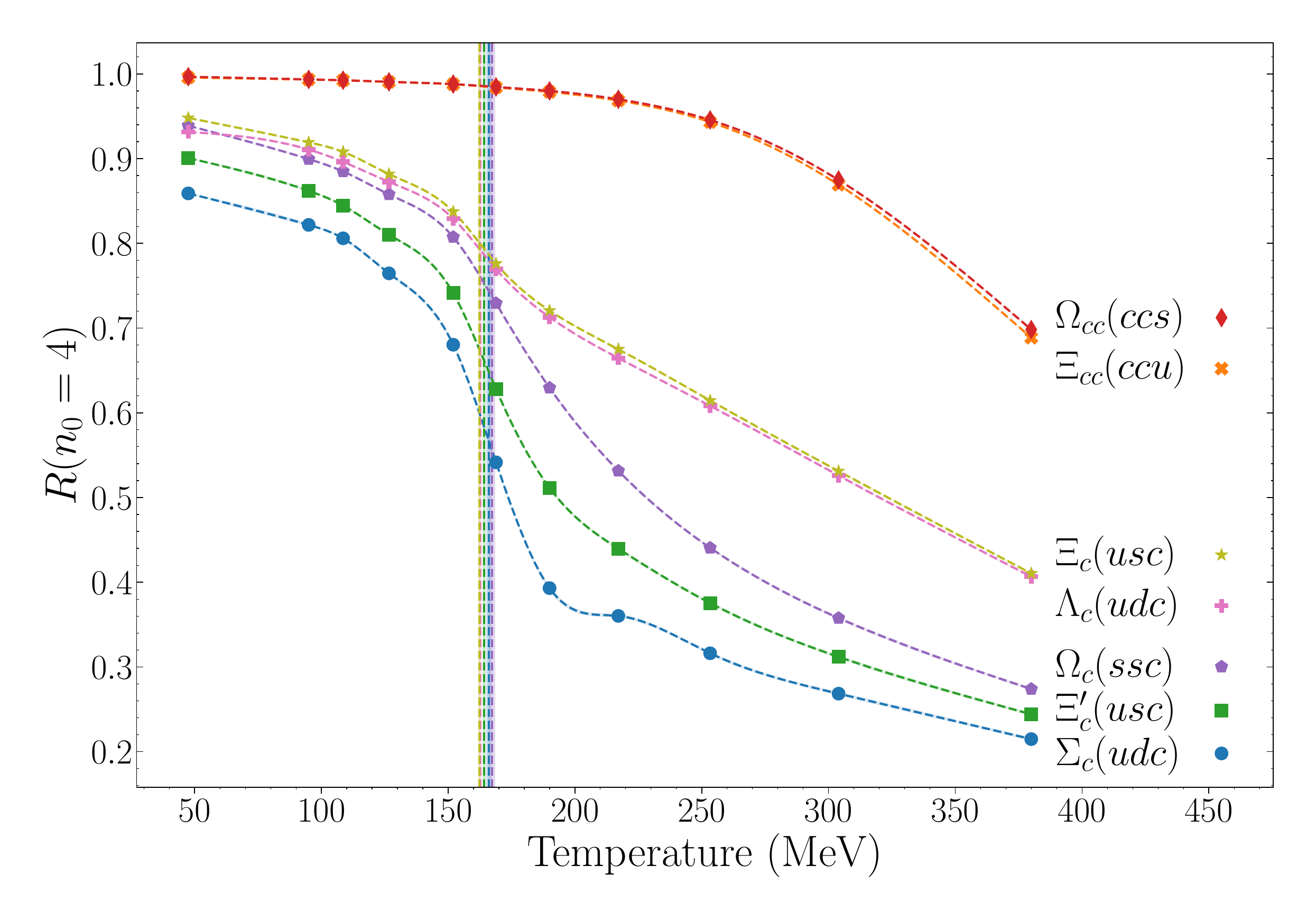}
  \caption{\label{fig:1/2RRatio}Crossover behaviour of the $R$ parameter of Eq.~(\ref{eqn:Rratio}) for $J=\frac{1}{2}$ baryons. The lines connecting the data points are cubic splines; the vertical lines indicate the inflection points for singly charmed baryons. No inflection point is found for the doubly-charmed baryons.}
\end{figure}

\Fig{fig:1/2RRatio} shows the $R$ parameter for the singly and doubly charmed  $J=\frac{1}{2}$ baryons  we considered. We observe that the values are indeed close to one at the lowest temperature, as expected. As the temperature increases, $R$ decreases for singly charmed baryons, indicating the sensitivity of the constituent light quarks to chiral symmetry restoration. Above the crossover temperature, the $R$ values are distinct from zero but continue to decrease. It is expected that at very high temperature the effect of the charm quark mass will vanish eventually, as $m_c/T \rightarrow 0$. For doubly charmed baryons, the $R$ parameter is much less sensitive to temperature and a reduction only sets in at higher temperatures.  It is interesting to note the ordering of the $R$ values at high temperature, in the QGP: $\Sigma_c$, $\Xi_c'$, and $\Omega_c$ belong to the SU(3) $=\mathbf{6}$ flavour multiplet, while $\Lambda_c$ and $\Xi_c$ belong to the SU(3)$=\mathbf{\bar 3}$ flavour multiplet. The doubly charmed baryons are significantly heavier.

The lines connecting the data points in \Fig{fig:1/2RRatio} are cubic splines fitted to the points. For singly charmed baryons, these splines have an inflection point indicated by the vertical lines and the inflection point temperatures are listed in \Tab{tab:inflect}. We note that within a few MeV the inflection point temperatures agree with the pseudocritical temperature obtained from the renormalised chiral condensate. This approximate agreement was observed earlier for the case of light and strange baryons, see Refs.~\cite{Aarts:2018glk,Aarts:2020vyb}. It is somewhat remarkable to note that even for singly charmed baryons the constituent light and strange quarks are sufficiently sensitive to chiral symmetry restoration to induce inflection points at these temperature values. On the other hand, the $R$ parameters for doubly charmed baryons do not show an inflection point in this temperature range.

\section{Conclusions}

In this work we presented a detailed study of the spin $1/2$ charmed baryons throughout the hadronic and into the deconfining phase. After determining the spectrum at the lowest temperature, the temperature dependence in each channel was investigated using ratios of thermal correlators with model or reconstructed correlators built on spectral content from the lowest temperature. This permits us to identify where thermal effects become important. It was found that the negative parity sector is sensitive to thermal effects at a much lower temperature than the positive parity sector. For doubly charmed, positive parity baryons we observed an approximate temperature independence of correlator ratios up to $T\sim 190$ MeV, the highest temperature we studied. The analysis of these ratios is robust as it does not rely on a particular Ansatz for spectral content at finite temperature.

Based upon the analysis of correlator ratios, we justify the use of multi-exponential fits to determine baryon masses in the cases where the spectral content appears to be largely unchanged. This allows us to extract positive parity masses at temperatures up to $T\sim 190$ MeV, albeit with a large uncertainty for singly charmed baryons, while the negative parity masses can only be determined up to $T\sim 127$ MeV. The masses of the positive parity ground states of doubly charmed baryons are approximately independent of temperature up to $T\sim 190$ MeV.

To study the effect of chiral symmetry restoration for charmed baryons, we investigated the $R$ parameter which encodes the difference between positive and negative parity correlators as a function of temperature. While parity doubling is not manifest --- as expected due to the large charm quark mass --- a crossover effect is observed for singly charmed baryons. Interestingly, the temperatures of the inflection points of this quantity are close to the pseudocritical temperature obtained from the renormalised chiral condensate, indicating that the constituent light and strange quarks are sufficiently sensitive to chiral symmetry restoration to induce this effect also in charmed correlators.

Future studies could examine the spectrum using either a more sophisticated operator basis or a lattice with more temporal points~\cite{Skullerud:2022yjr}. Both these methods are under investigation. Alternatively one could use spectral function reconstruction methods to directly examine the mass and width of the ground states as a function of temperature to determine when thermal effects become significant.

\begin{table}[t]
  \caption{\label{tab:inflect}Temperature of the inflection points (in MeV) of the $R$ parameter (\ref{eqn:Rratio}) for singly charmed baryons presented in Figure~\ref{fig:1/2RRatio}, categorised by strangeness (S) and charm (C). Uncertainties are statistical only. For reference, the pseudocritical temperature from the renormalised chiral condensate is $\Tpc= 167(2)(1)$ MeV.}
  \centering
  \begin{tabular}{cc|ll}
    S       & C  $\;$   & \multicolumn{2}{c}{Inflection point (MeV)}                                            \\ \hline
    $0$     & $1$ $\;$  & \multicolumn{1}{l|}{$\;$ $\Sigma_{c}\rb{udc}$       $166.1(1.0)$ $\;$ } & $\;$ $\Lambda_{c}\rb{udc}$ $162.6(5)$ \\
    $-1$    & $1$ $\;$  & \multicolumn{1}{l|}{$\;$ $\Xi^{\prime}_{c}\rb{usc}$ $164.2(6)$}    & $\;$ $\Xi_{c}\rb{usc}$     $162.3(4)$  \\
    $-2$    & $1$ $\;$  & \multicolumn{1}{l|}{$\;$ $\Omega_{c}\rb{ssc}$       $167.2(1.3)$}  &        
  \end{tabular}
\end{table}


\begin{figure*}[t]
  \centering
  \includegraphics[width=0.42\linewidth]{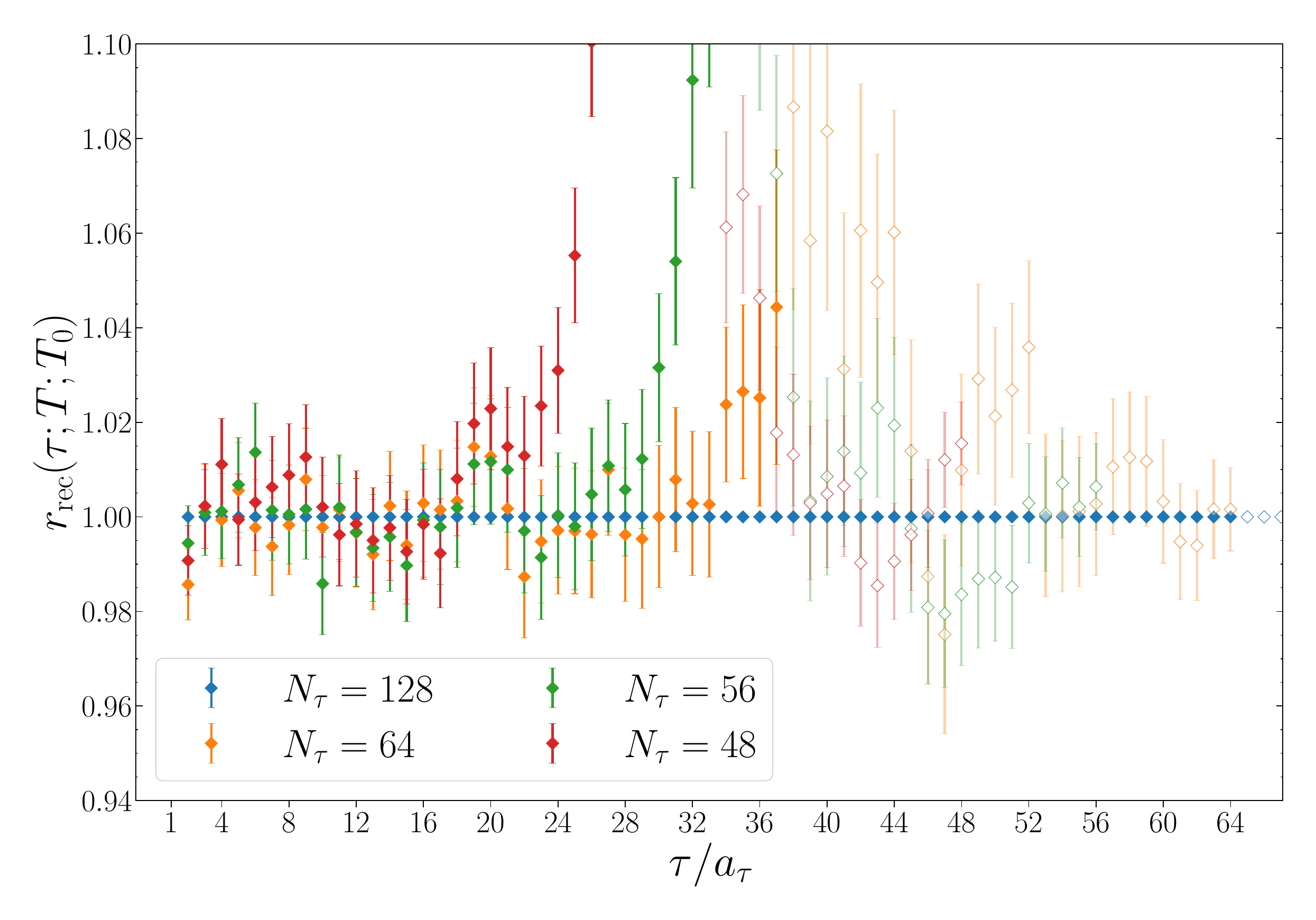}
  \includegraphics[width=0.42\linewidth]{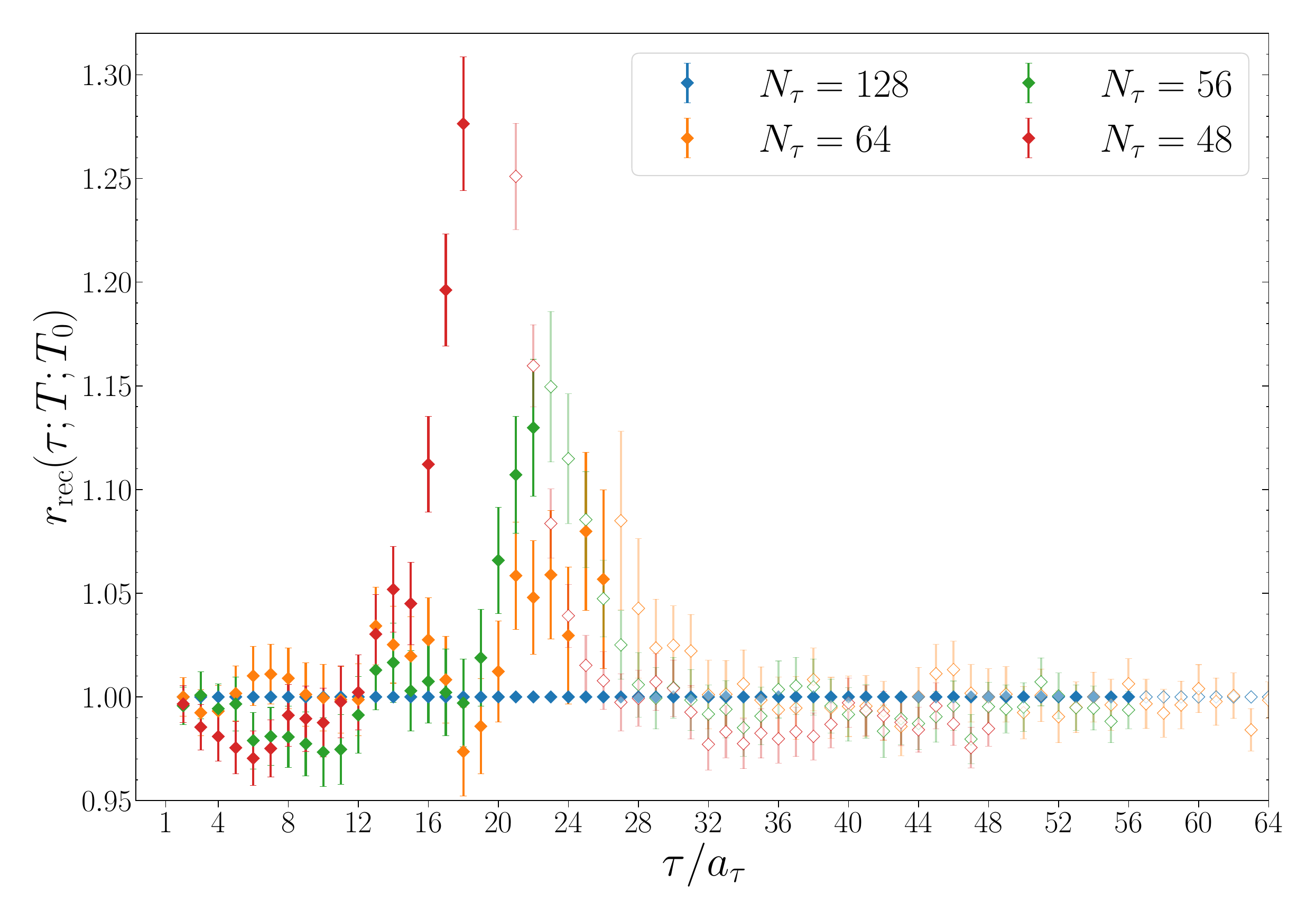}
  \\
  \vspace{-0.0053\textheight}
  \includegraphics[width=0.42\linewidth]{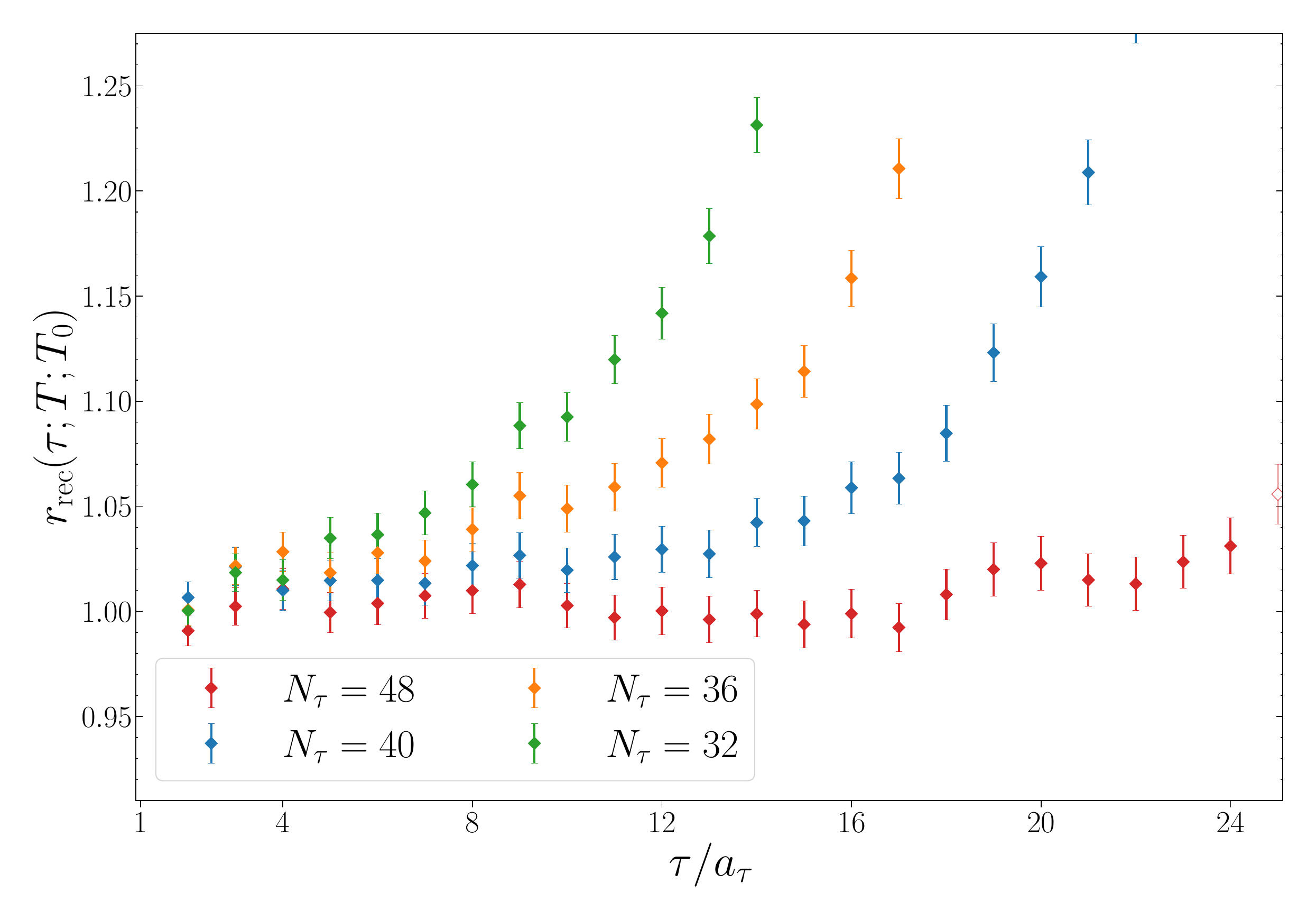}
  \includegraphics[width=0.42\linewidth]{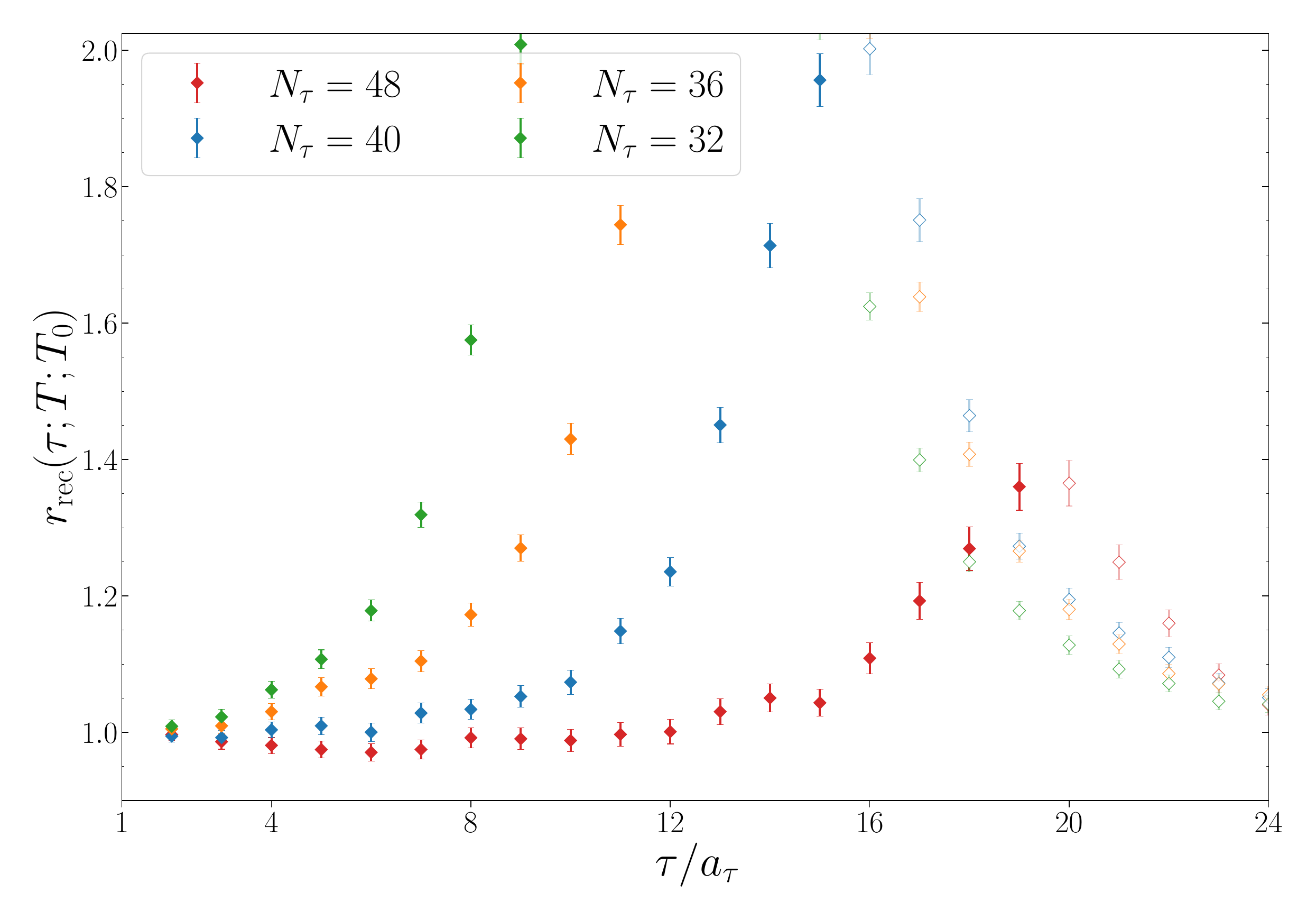}
  \caption{Ratio with the reconstructed correlator as in \Fig{fig:sigma12_3fl_udc_single_recon}, in the $\Omega_{cc}\rb{ccs}$ channel.}
  \label{fig:doublet_2fl_ccs_single_recon}
\end{figure*}

\vspace{-0.2cm}
\section*{Software and Data}

Correlation functions were generated using \textsc{openQCD-hadspec}~\cite{glesaaen_jonas_2018_2217028}, an extension to \textsc{openQCD-FASTSUM}~\cite{glesaaen_jonas_rylund_2018_2216356} for correlation functions. \textsc{openQCD-FASTSUM} was used for ensemble generation~\cite{Aarts:2020vyb,Aarts:2022krz}.

The analysis in this work makes ample use of the \textsc{Py-thon} packages \textsc{gvar}  \cite{peter_lepage_2022_7315961} and \textsc{lsqfit} \cite{peter_lepage_2021_5777652}. Additional data analysis tools included \textsc{matplotlib} \cite{Hunter:2007,thomas_a_caswell_2022_6513224} and \textsc{numpy} \cite{harris2020array}. The dataset and scripts used for this paper can be found at~\cite{aarts_gert_2023_8273591}.

\section*{Authors' Contributions}

\begin{itemize}
\item Bignell: Data analysis, plot generation and primary manuscript production.
\item J\"ager: Data production and preliminary data analysis.
\item Aarts: Initial proposal \cite{Aarts:2018haw}, development of the reconstructed correlator for baryons and draft of the \\manuscript.
\item Allton, Anwar, Burns, Skullerud: Contributions to analysis and physics interpretations of results, and draft of the manuscript.
\end{itemize}

\begin{acknowledgement}
G.A., C.A., R.B.\ and T.J.B.\ are grateful for support via STFC grant ST/T000813/1. 
M.N.A.\ acknowledges support from The Royal Society Newton International Fellowship.
This work used the DiRAC Extreme Scaling service at the University of Edinburgh, operated by the Edinburgh Parallel Computing Centre and the DiRAC Data Intensive service operated by the University of Leicester IT Services on behalf of the STFC DiRAC HPC Facility (www.dirac.ac.uk). This equipment was funded by BEIS capital funding via STFC capital grants ST/R00238X/1, ST/K000373/1 and ST/R002363/1 and STFC DiRAC Operations grants ST/R001006/1 and ST/R001014/1. DiRAC is part of the UK National e-Infrastructure. We acknowledge the support of the Swansea Academy for Advanced Computing, the Supercomputing Wales project, which is part-funded by the European Regional Development Fund (ERDF) via Welsh Government, and the University of Southern Denmark and ICHEC, Ireland for use of computing facilities. This work was performed using PRACE resources at Cineca (Italy), CEA (France) and Stuttgart (Germany) via grants 2015133079, \\2018194714, 2019214714 and 2020214714. 
\end{acknowledgement}

\appendix

\section{Ratios in other channels}
\label{sec:app}

\begin{figure*}[!p]
  \centering
  \includegraphics[width=0.42\linewidth]{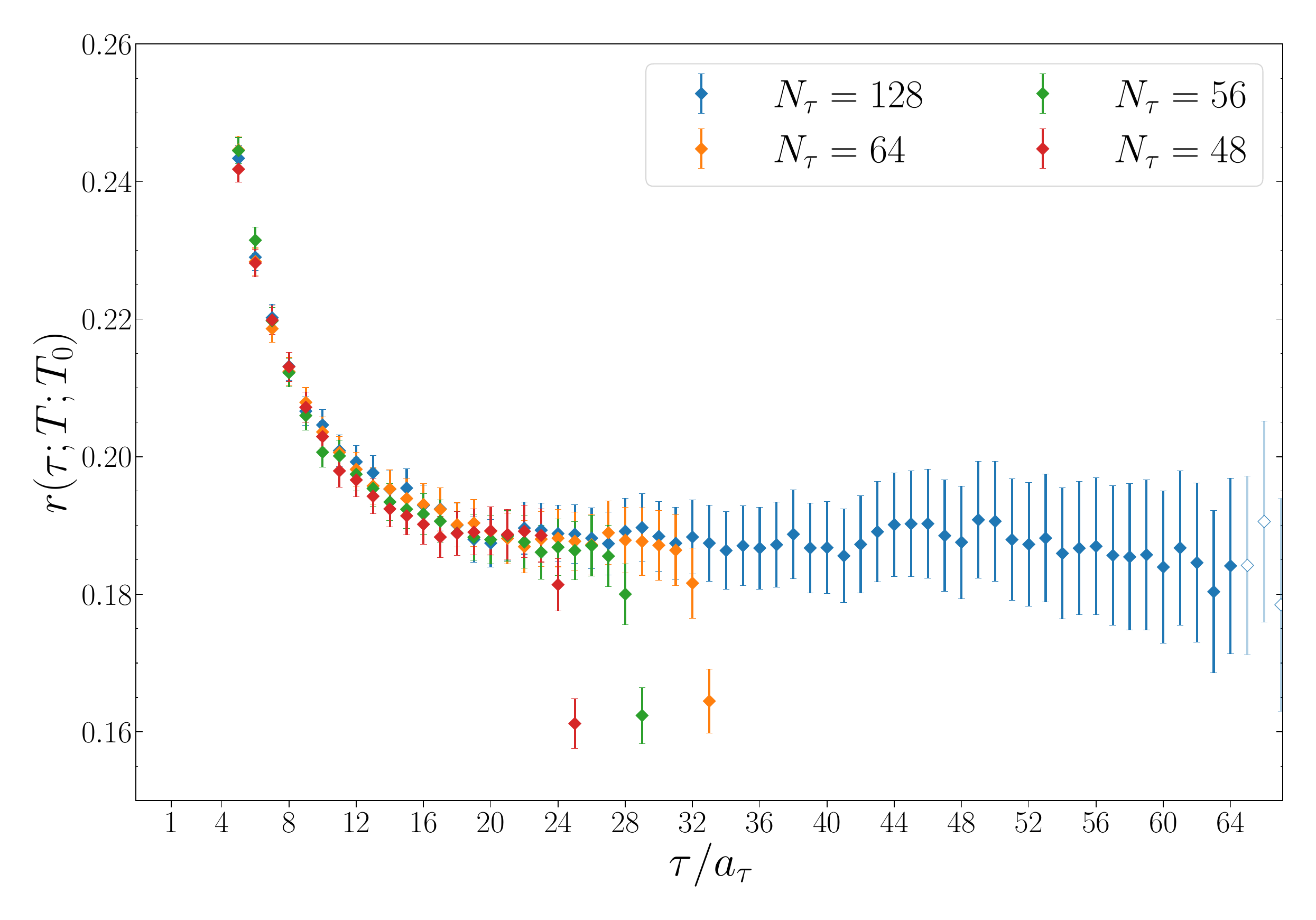}
  \includegraphics[width=0.42\linewidth]{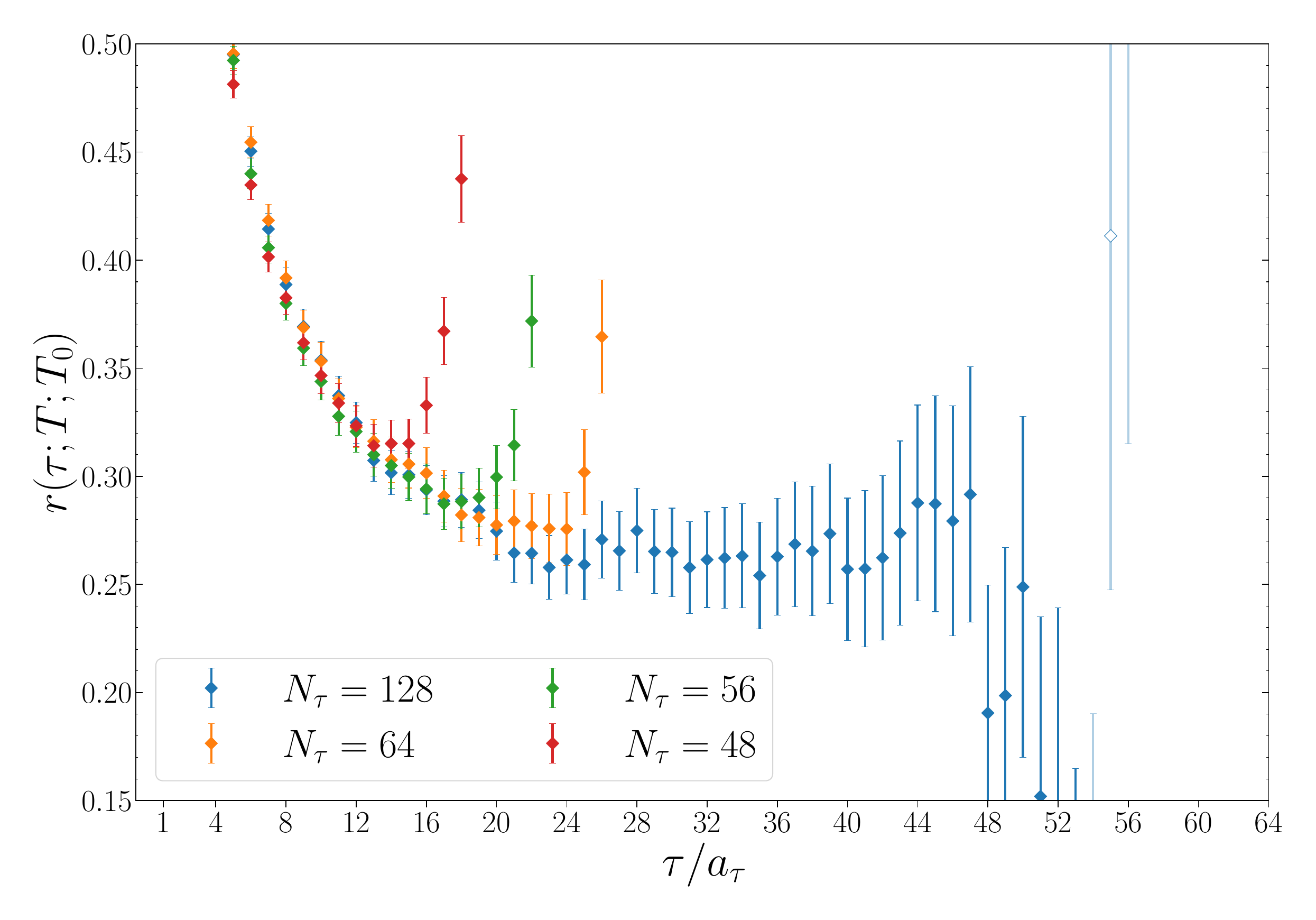}
  \\
  \vspace{-0.0053\textheight}
  \includegraphics[width=0.42\linewidth]{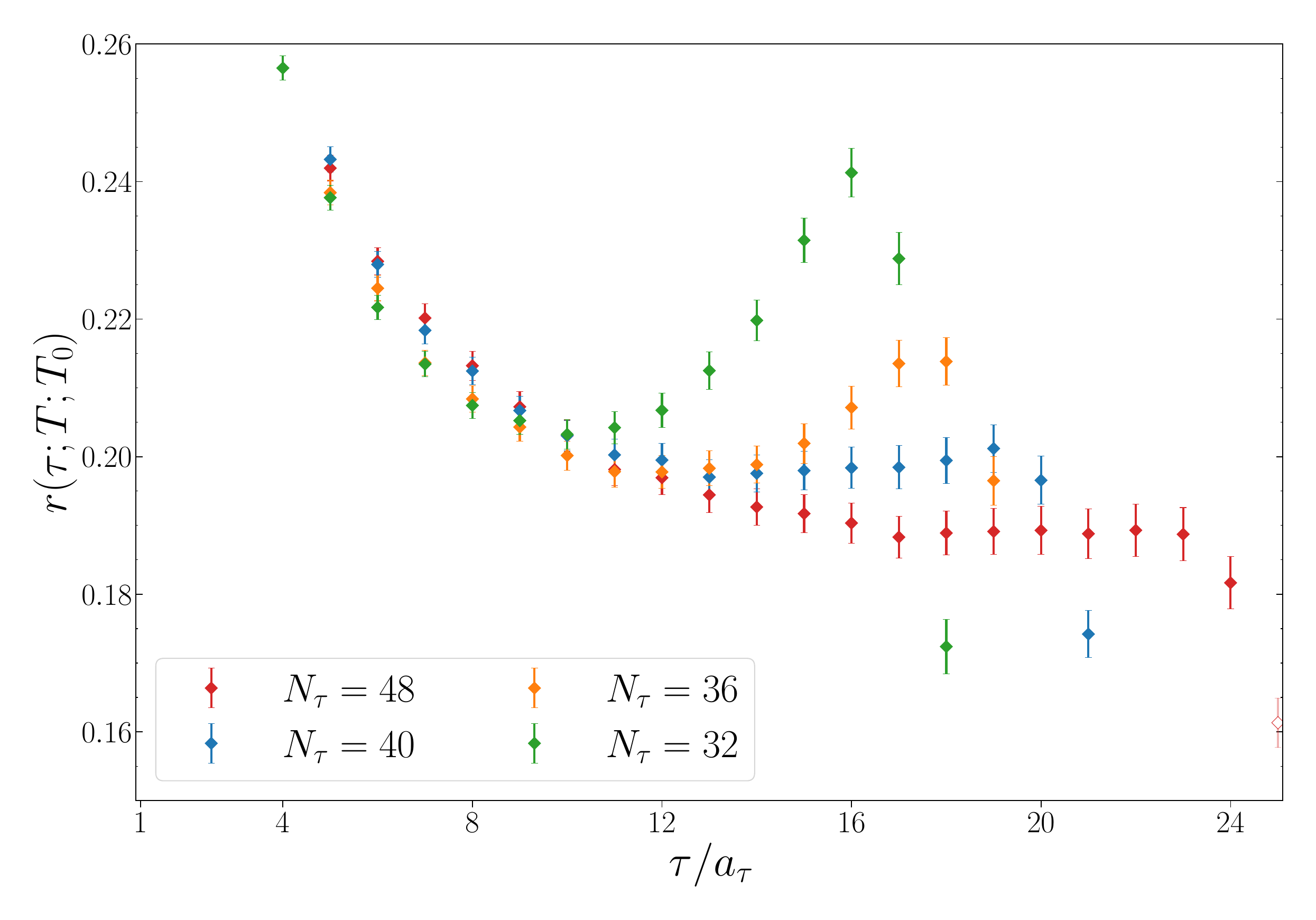}
  \includegraphics[width=0.42\linewidth]{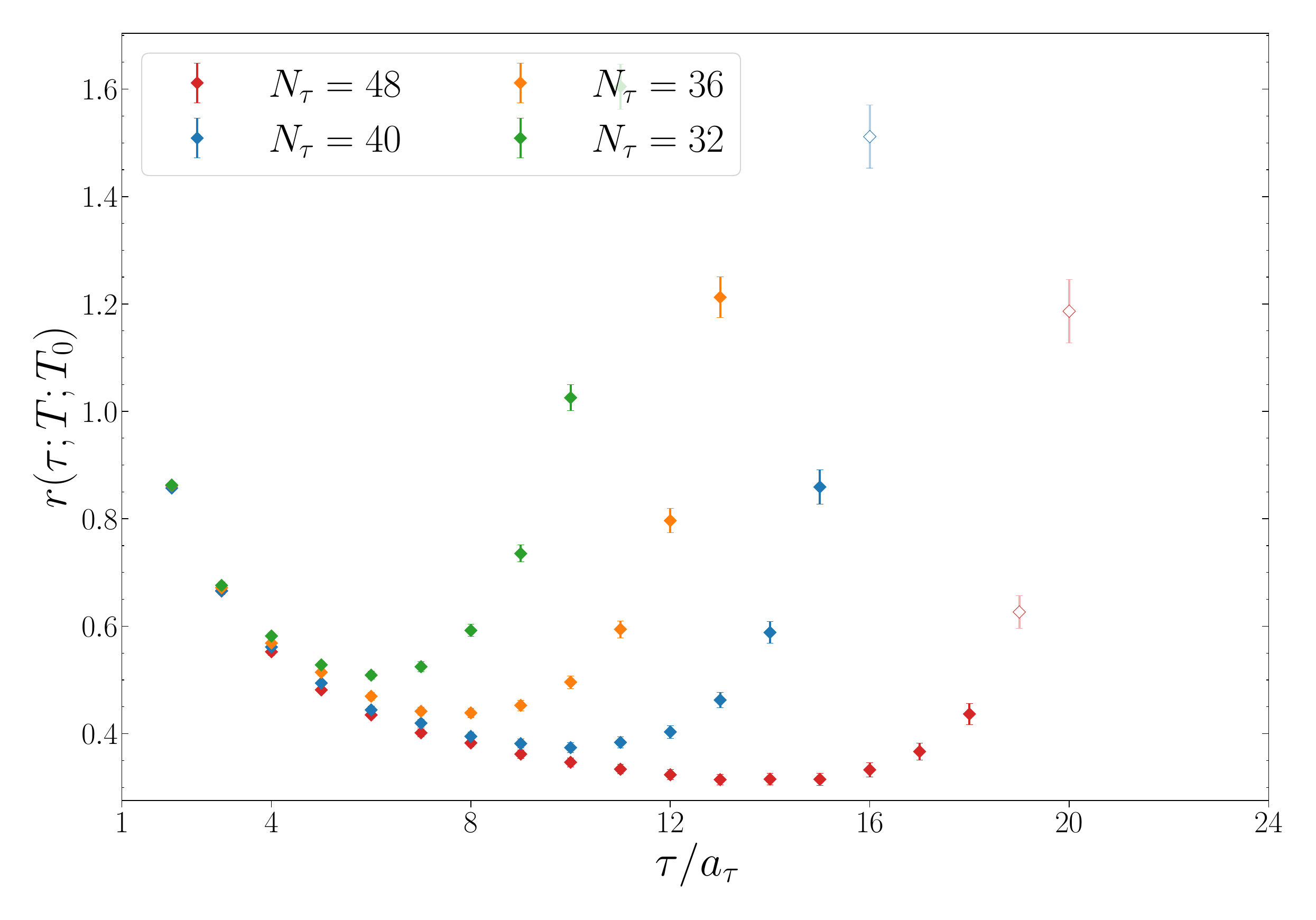}
  \caption{Single ratio with the model correlator as in \Fig{fig:sigma12_3fl_udc_single}, in the $\Xi_{cc}\rb{ccu}$ channel.}
  \label{fig:doublet_2fl_ccu_single}

\vspace{0.5cm}

  \centering
  \includegraphics[width=0.42\linewidth]{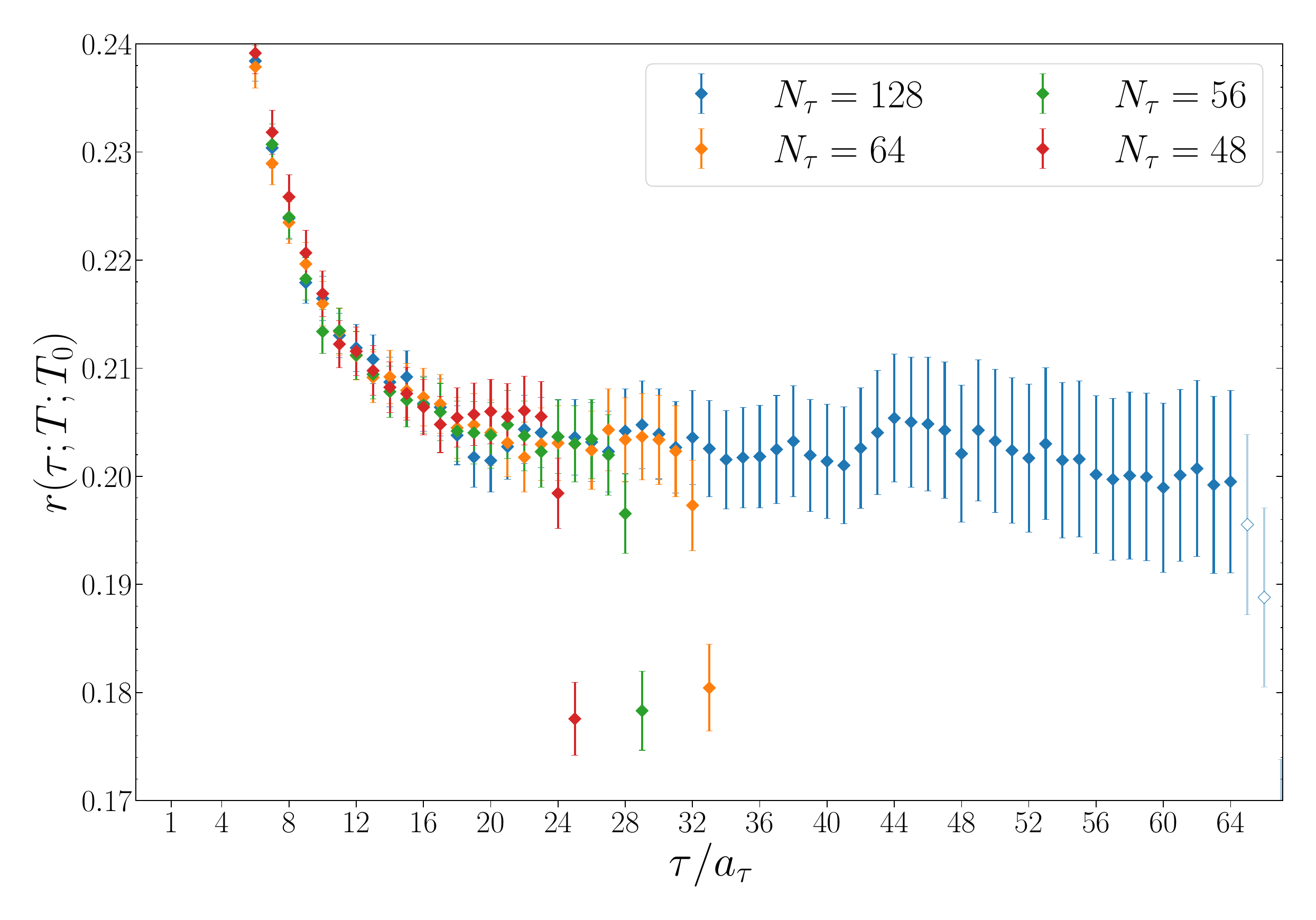}
  \includegraphics[width=0.42\linewidth]{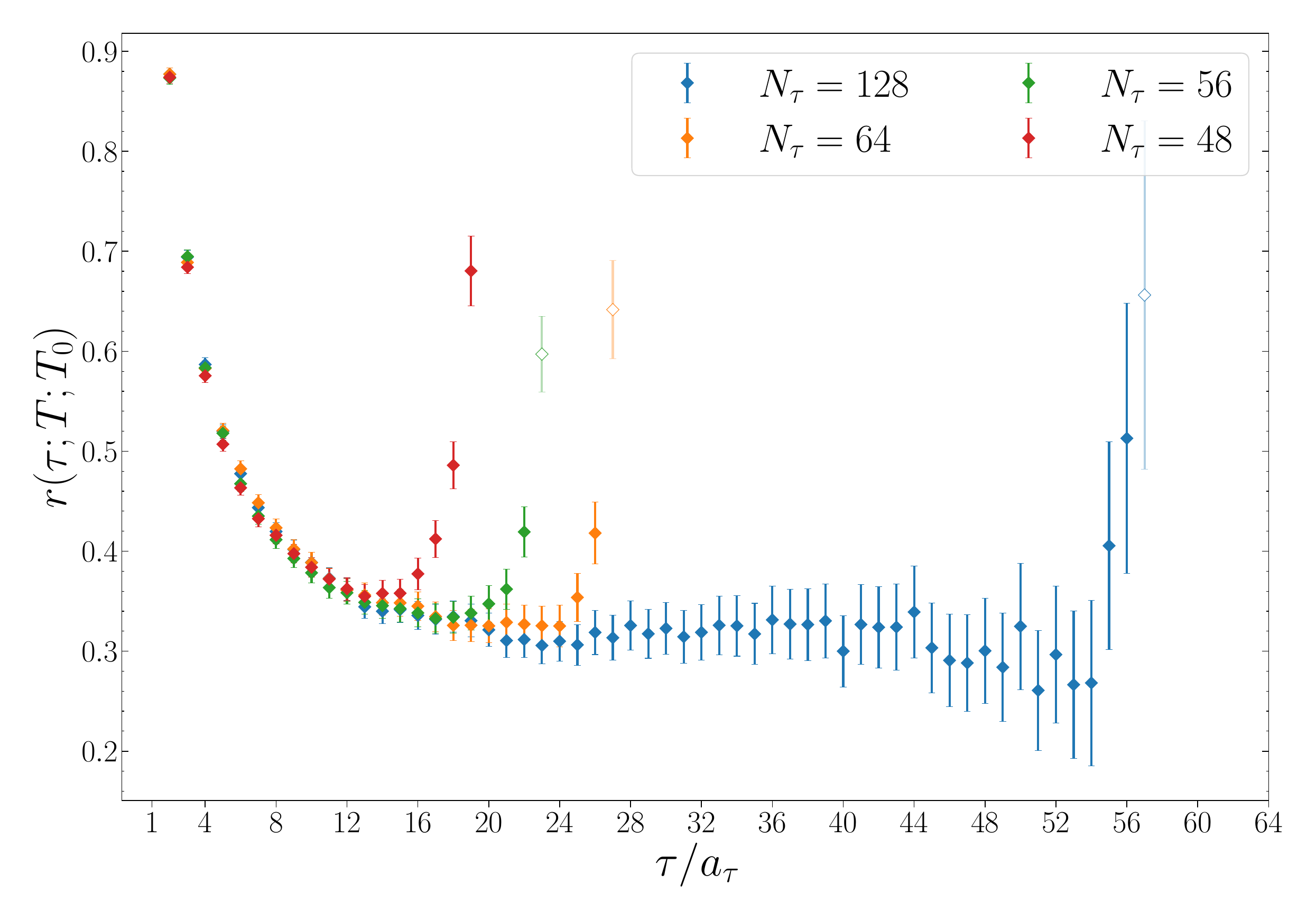}
  \\
  \vspace{-0.0053\textheight}
  \includegraphics[width=0.42\linewidth]{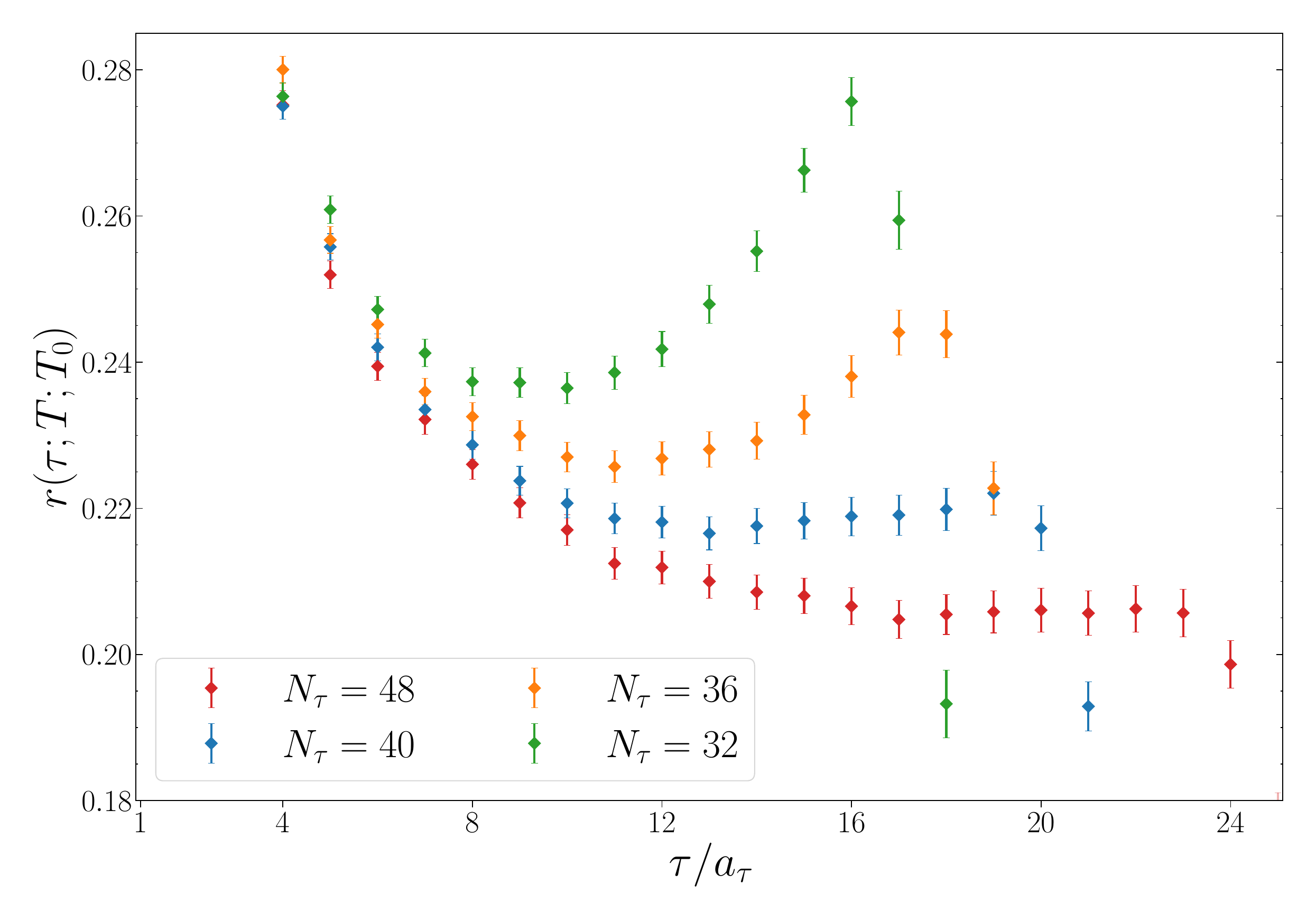}
  \includegraphics[width=0.42\linewidth]{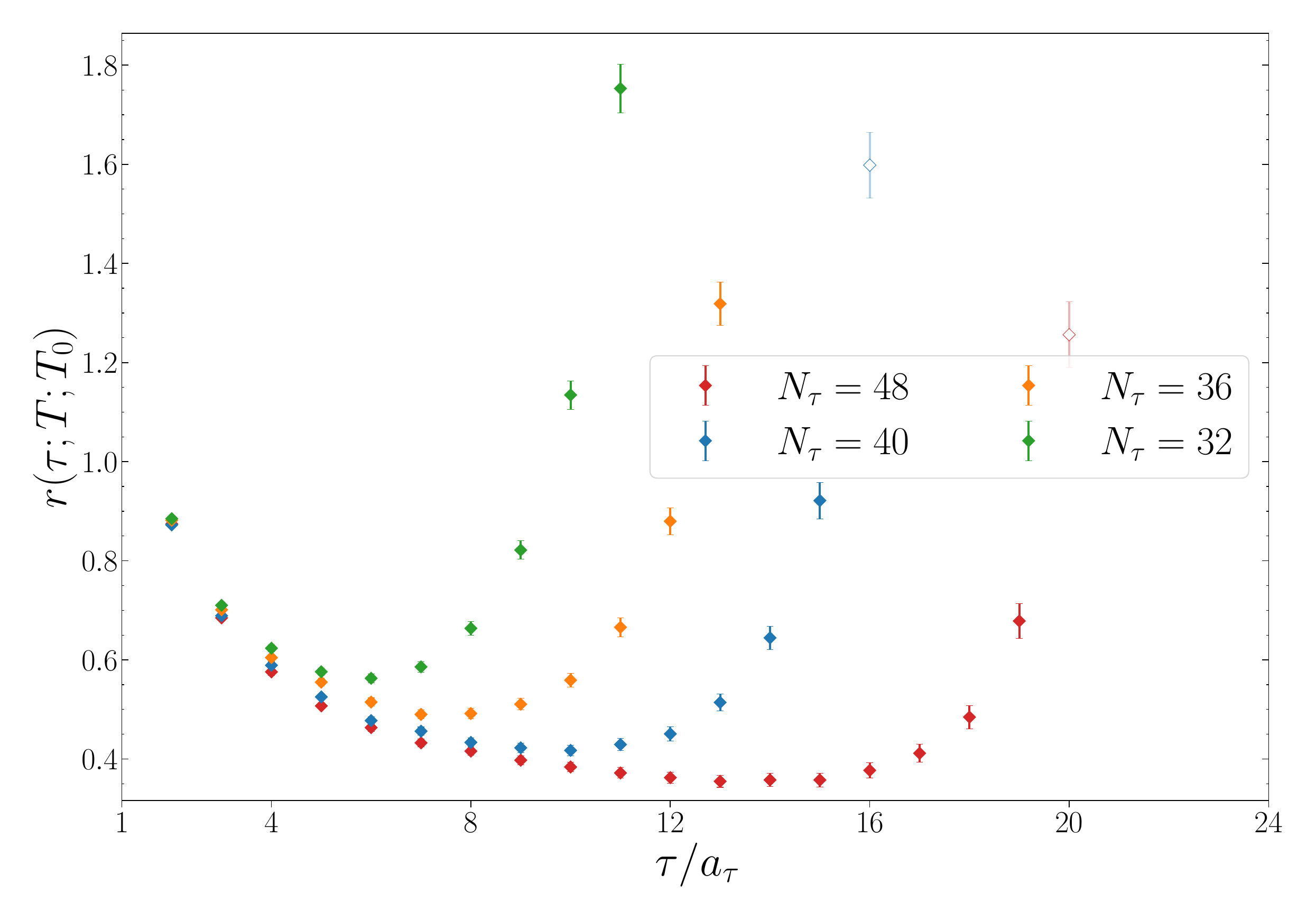}
  \caption{Single ratio with the model correlator as in \Fig{fig:sigma12_3fl_udc_single}, in the $\Omega_{cc}\rb{ccs}$ channel.}
  \label{fig:doublet_2fl_ccs_single}
\end{figure*}

\begin{figure*}[!p]
  \centering
  \includegraphics[width=0.42\linewidth]{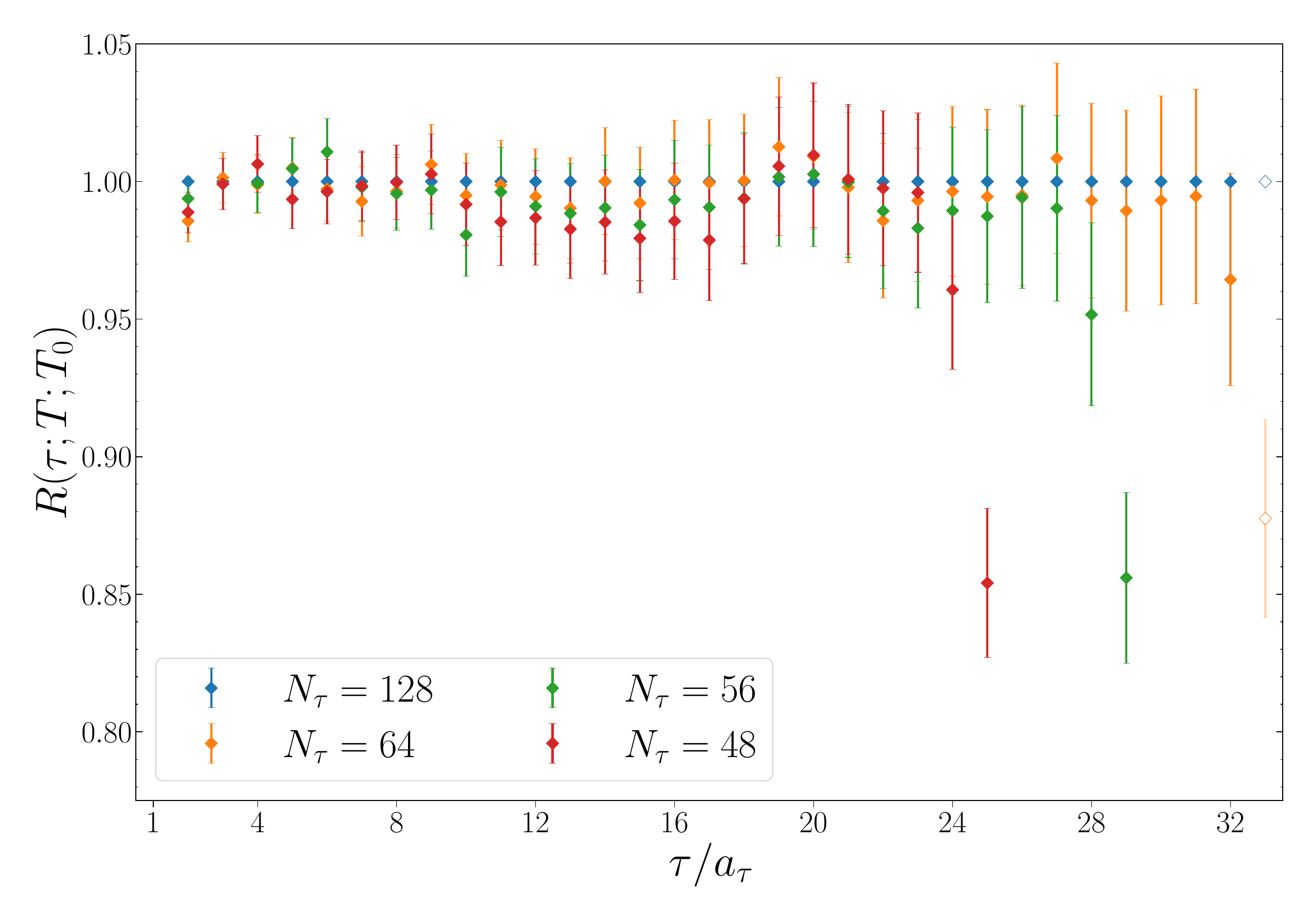}
  \includegraphics[width=0.42\linewidth]{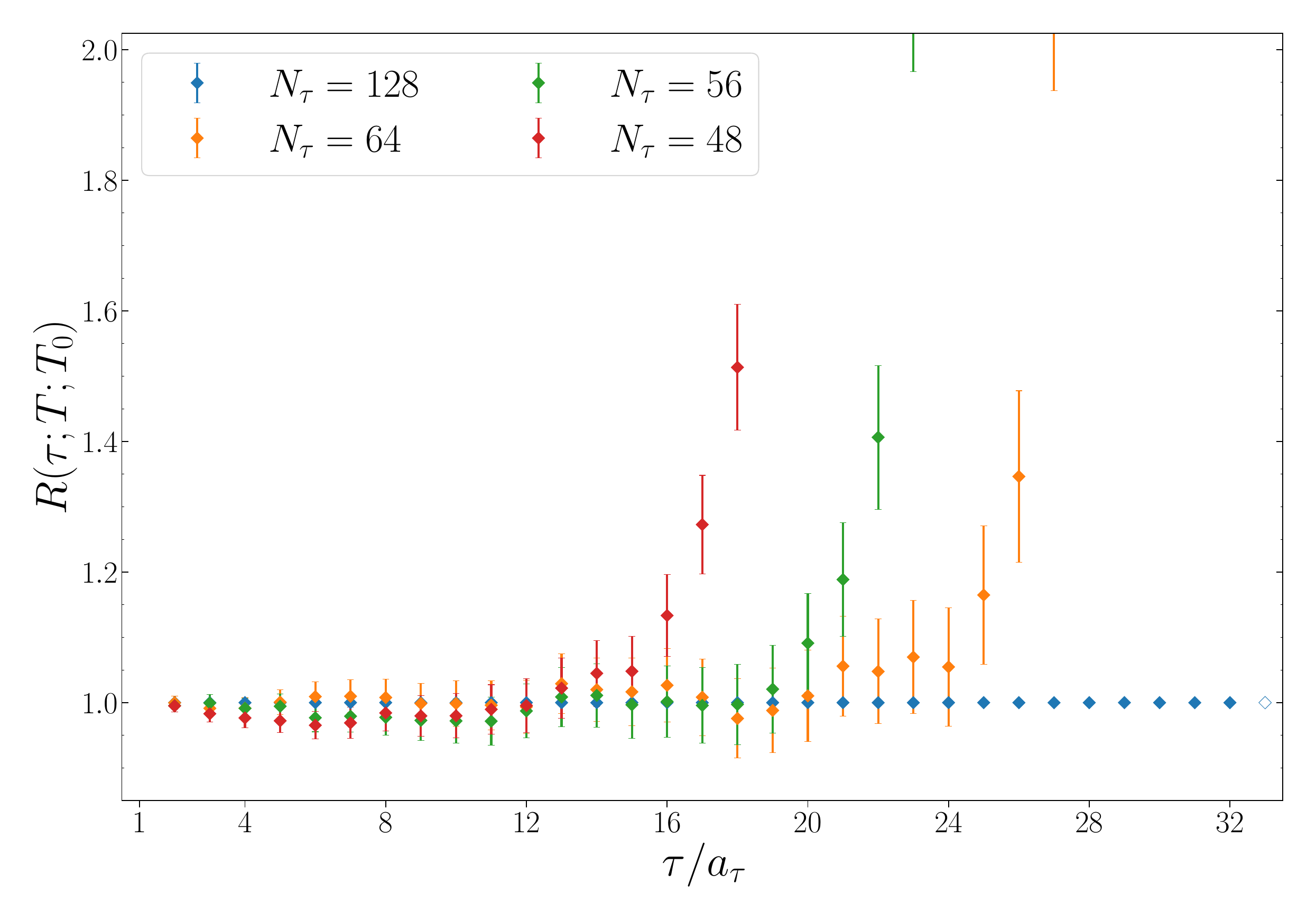}
  \\
  \vspace{-0.0053\textheight}
  \includegraphics[width=0.42\linewidth]{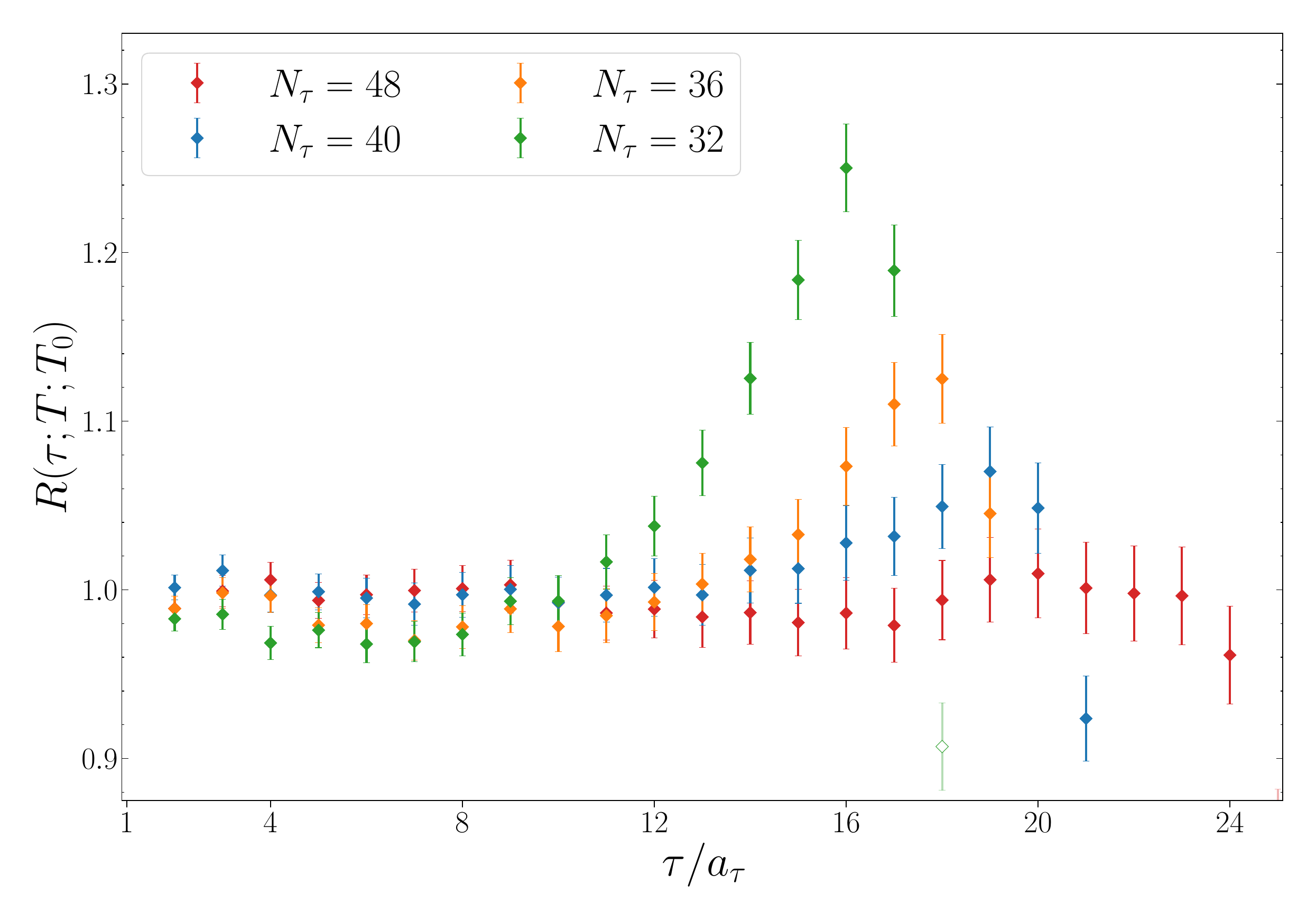}
  \includegraphics[width=0.42\linewidth]{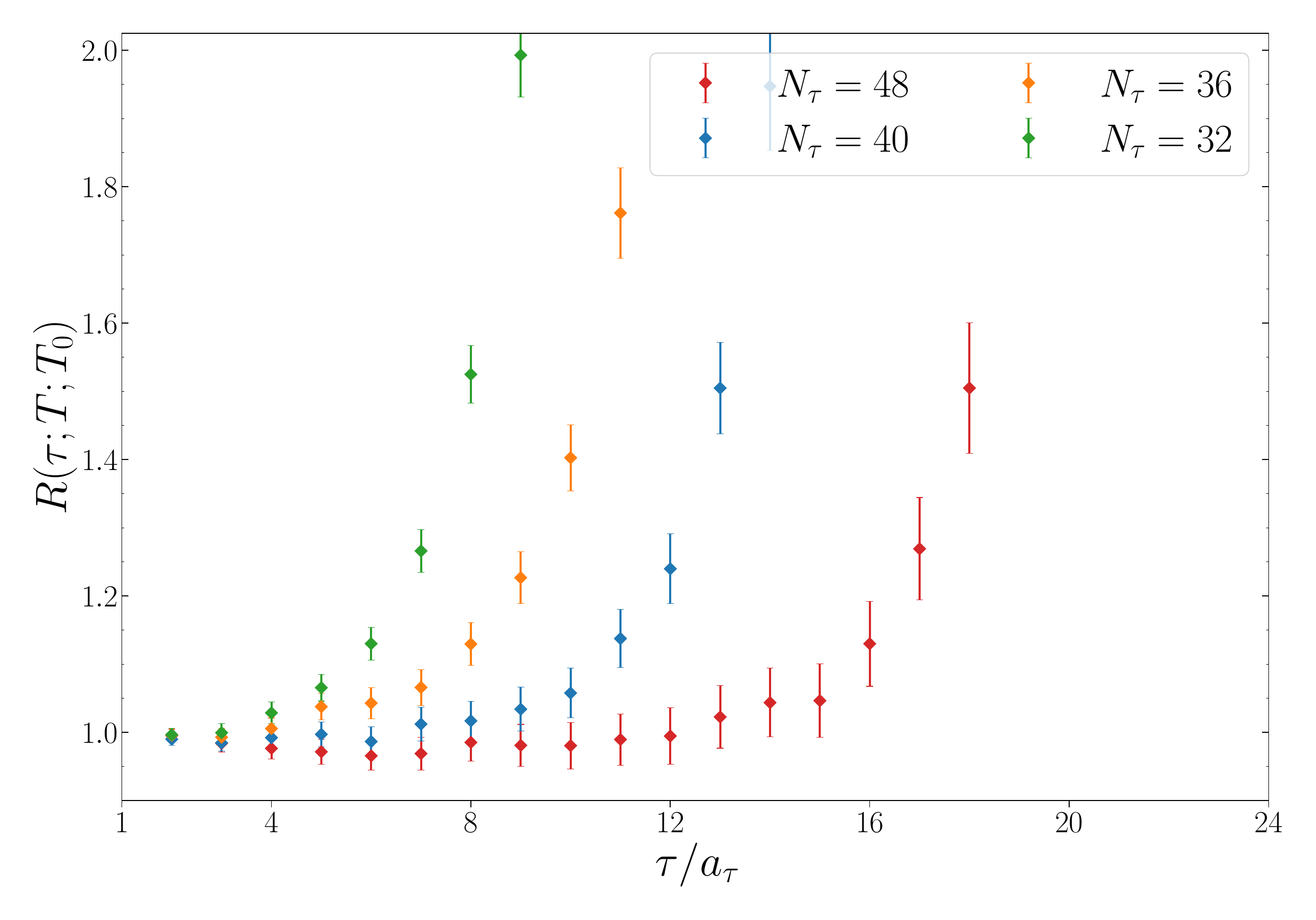}
  \caption{Double ratio as in \Fig{fig:sigma12_3fl_udc_double}, in the $\Xi_{cc}\rb{ccu}$ channel.}
  \label{fig:doublet_2fl_ccu_double}

\vspace{0.5cm}

  \centering
  \includegraphics[width=0.42\linewidth]{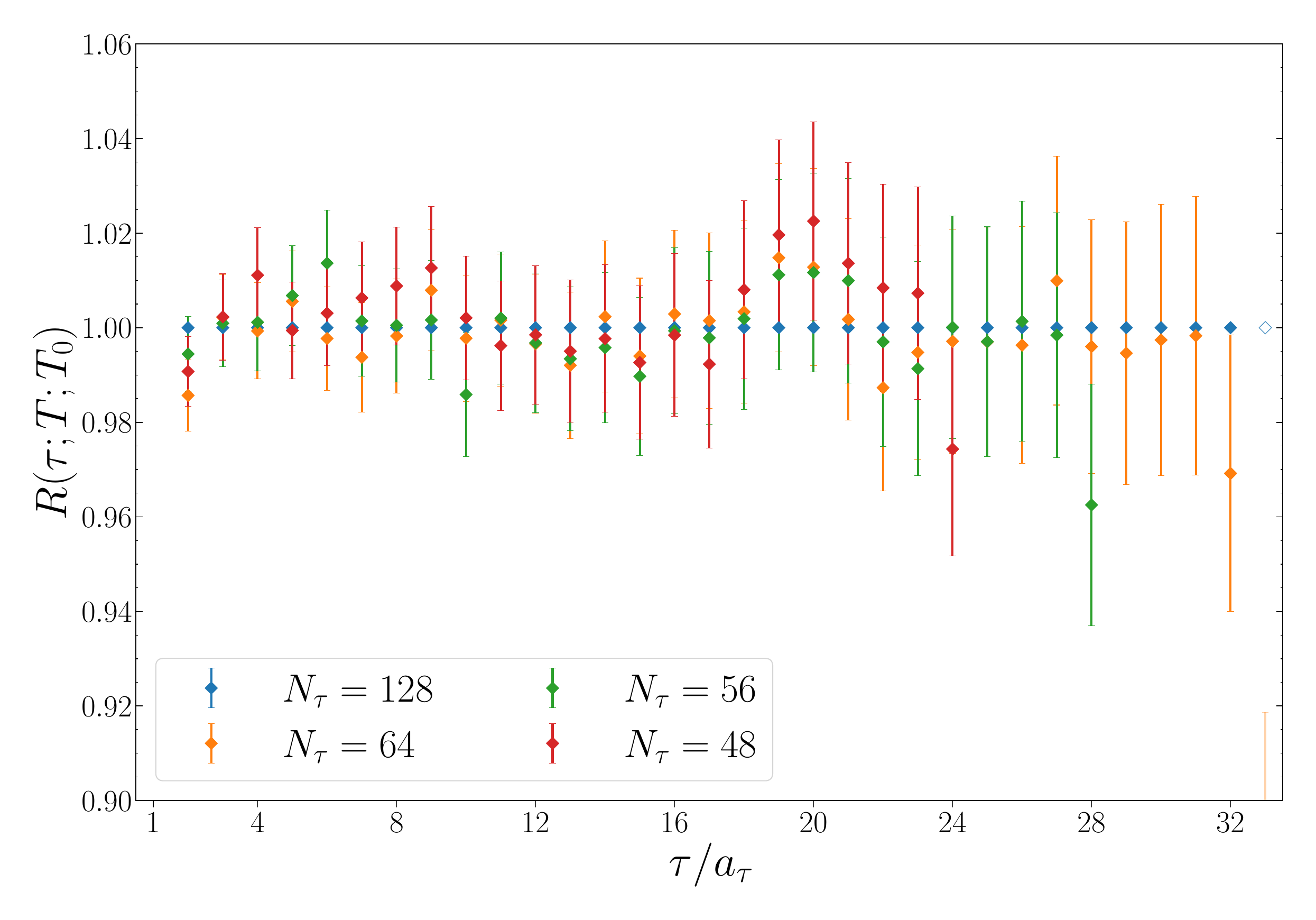}
  \includegraphics[width=0.42\linewidth]{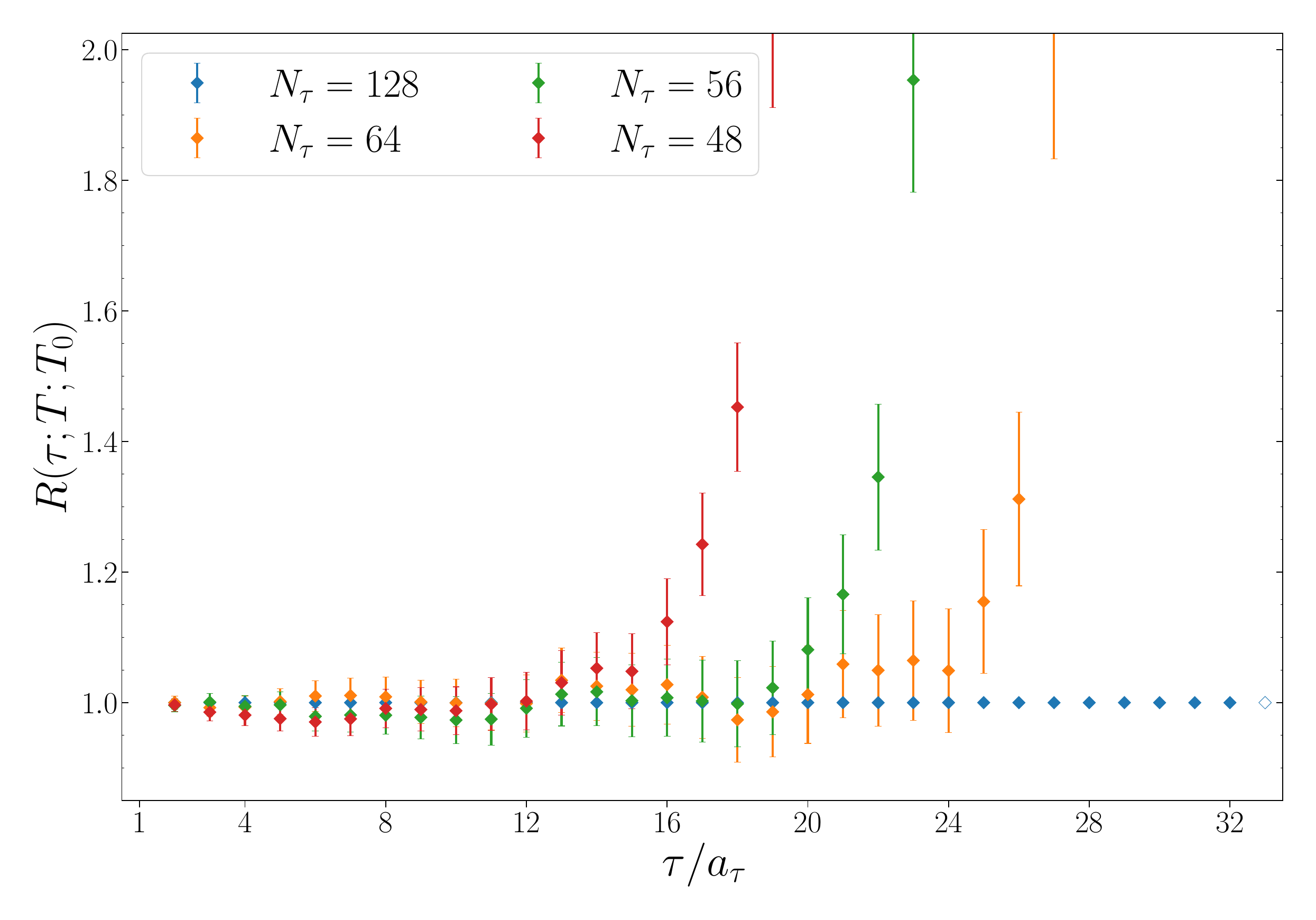}
  \\
  \vspace{-0.0053\textheight}
  \includegraphics[width=0.42\linewidth]{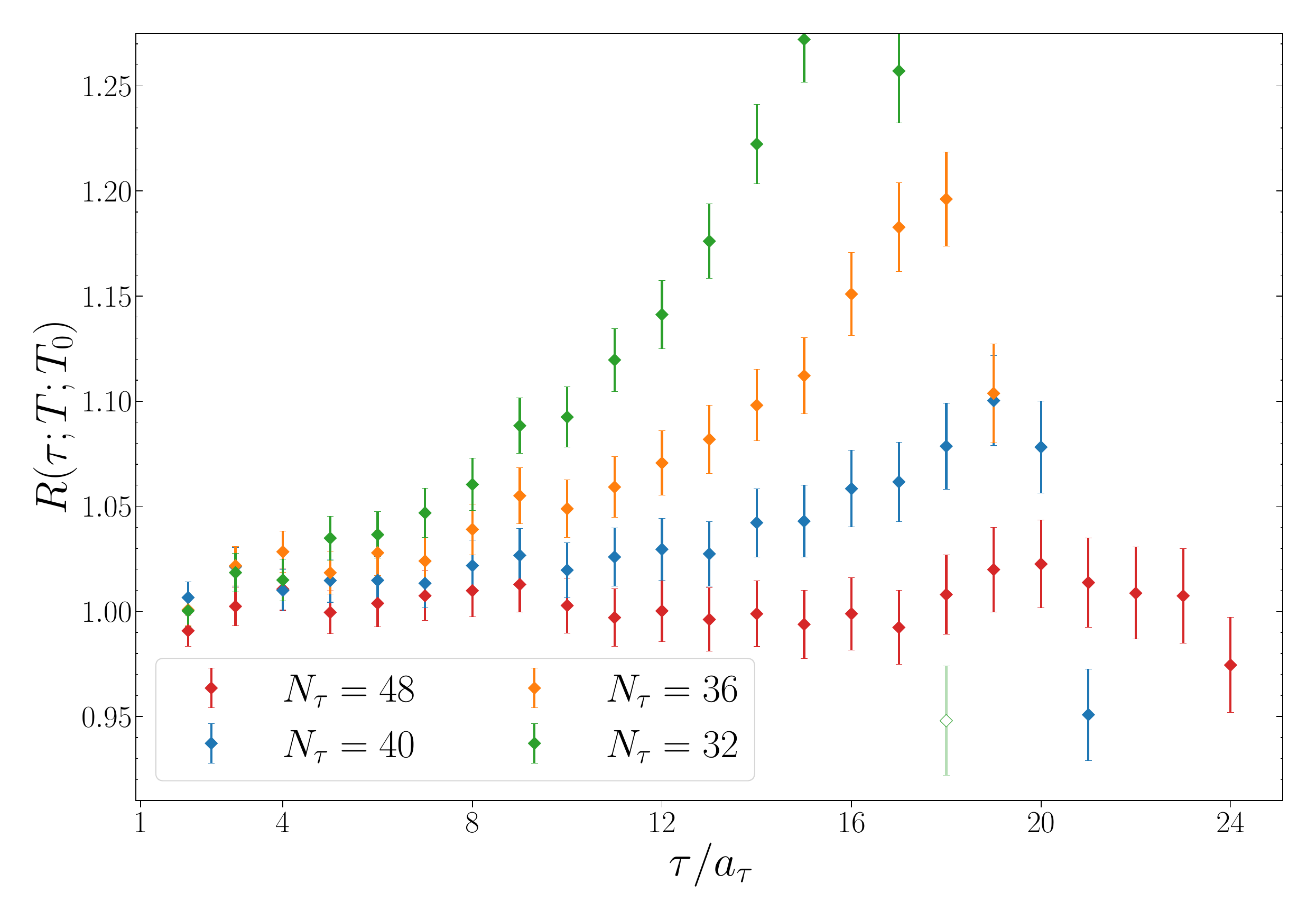}
  \includegraphics[width=0.42\linewidth]{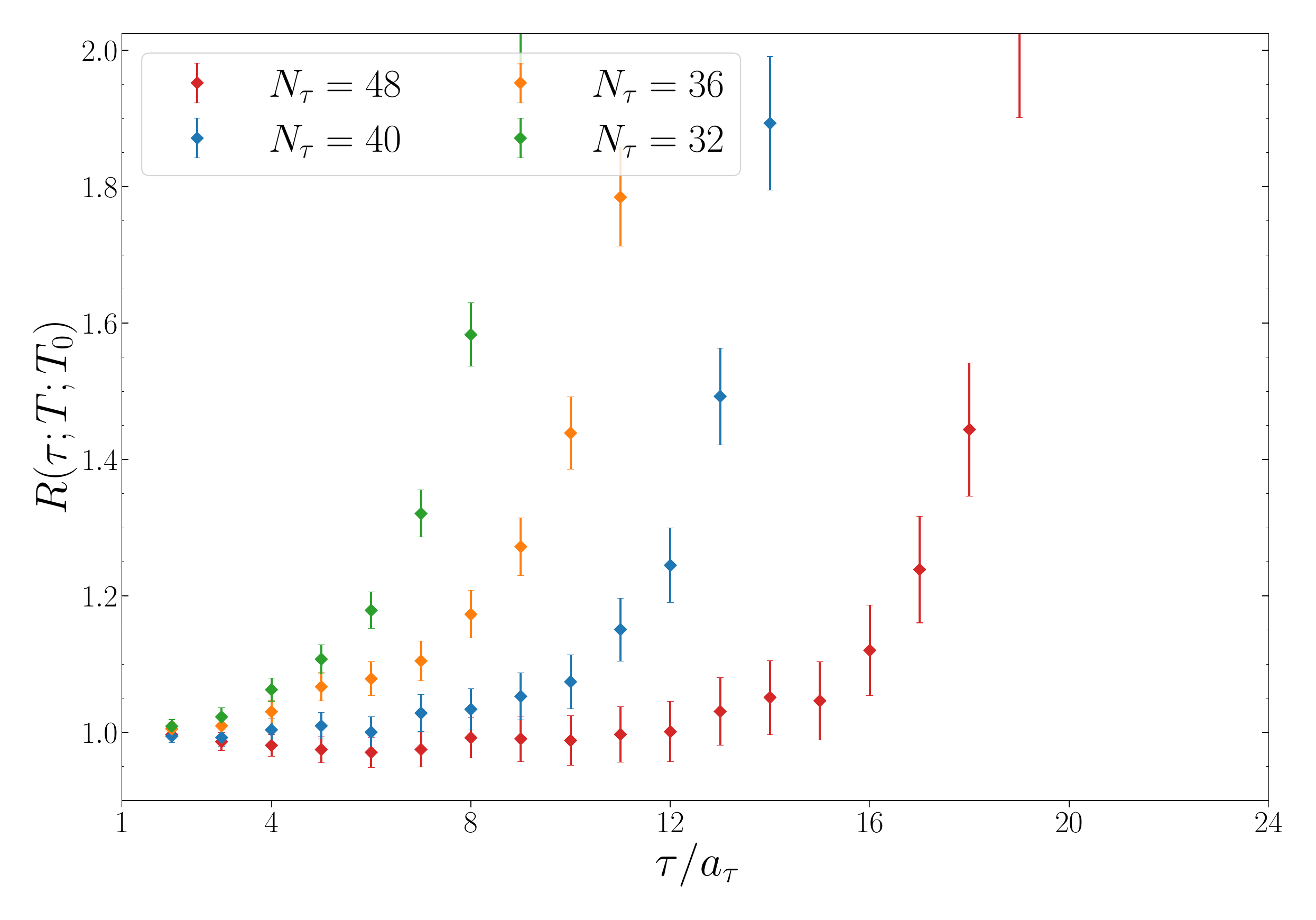} 
  \caption{Double ratio as in \Fig{fig:sigma12_3fl_udc_double} in the $\Omega_{cc}\rb{ccs}$ channel.}
  \label{fig:doublet_2fl_ccs_double}
\end{figure*}

For comparison, we present the ratios of thermal correlators with reconstructed and model correlators in the $\Xi_{cc}\rb{ccu}$ and the $\Omega_{cc}\rb{ccs}$ channels in Figs.~\ \ref{fig:doublet_2fl_ccs_single_recon}-\ref{fig:doublet_2fl_ccs_double}. Details are as in \Sec{sec:ratios}.


\bibliographystyle{JHEP_arXiv}
\typeout{}
\bibliography{charm}
\end{document}